\newcommand{\version}{November 29, 2000}

\documentclass[12pt]{article}

\usepackage{a4,amsthm,amsfonts,latexsym,amssymb}

{\catcode `\@=11 \global\let\AddToReset=\@addtoreset}
\AddToReset{equation}{section}

\swapnumbers
\theoremstyle{plain}
\newtheorem{thm}{THEOREM}[section]
\newtheorem{cl}[thm]{COROLLARY}
\newtheorem{lem}[thm]{LEMMA}
\newtheorem{prop}[thm]{PROPOSITION}
\theoremstyle{definition}
\newtheorem{rem}[thm]{Remark}

\newcommand{\beq}{\begin{equation}}
\newcommand{\eeq}{\end{equation}}
\def\beqa{\begin{eqnarray}}
\def\eeqa{\end{eqnarray}}

\newcommand{\R}{{\mathbb R}}
\newcommand{\C}{{\mathbb C}}
\newcommand{\N}{{\mathbb N}}

\newcommand{\eps}{\varepsilon}
\newcommand{\A}{{\bf A}}

\newcommand{\x}{{\bf x}}

\newcommand{\xperp}{\x_\perp}

\newcommand{\Tr}{{\rm Tr}}

\newcommand{\Ed}{E^{\rm DSTF}}

\newcommand{\E}{{\mathcal E}^{\rm DSTF}}
\newcommand{\Ec}{E_{\rm conf}}

\newcommand{\Eo}{{\mathcal E}^{1\rm D}_{B,Z}}

\newcommand{\vph}{\varphi^{(m)}_{\rm eff}(z)}
\newcommand{\rmd}{d}
\newcommand{\ED}{{\mathcal E}^{\rm DDM}_{B,Z}}
\newcommand{\Eddm}{E^{\rm DDM}}
\date{\small\version}

\begin{document}
\markboth{\scriptsize{H \version}}{\scriptsize{H \version}}
\title{\bf{Gradient corrections for semiclassical theories of atoms in \\ strong magnetic fields}}
\author{\vspace{5pt} Christian Hainzl\\
\vspace{-4pt}\small{Institut f\"ur Theoretische Physik, Universit\"at Wien}\\
\small{Boltzmanngasse 5, A-1090 Vienna, Austria}\\
\small{E-Mail:\,\texttt{hainzl@thp.univie.ac.at}}}

\maketitle

\begin{abstract}
This paper is divided into two parts. In the first one the von
Weizs\"acker term is introduced to the Magnetic TF theory
and the resulting MTFW functional is mathematically analyzed. In particular,
it is shown that the von Weizs\"acker term produces the Scott
correction up to magnetic fields of order $B \ll Z^2$, in accordance
with a result of V. Ivrii on the quantum mechanical ground state energy.

The second part is dedicated to gradient corrections for
semiclassical theories of atoms restricted to electrons in the lowest Landau
band. We consider modifications of the Thomas-Fermi theory for
strong magnetic fields (STF), i.e. for $B \ll Z^3$. The main
modification consists in replacing the integration over the
variables perpendicular to the field by an expansion in angular
momentum eigenfunctions in the lowest Landau band. This leads to
a functional (DSTF) depending on a sequence of one-dimensional
densities. For a one-dimensional Fermi gas the analogue of a
Weizs\"acker correction has a negative sign and we discuss the
corresponding modification of the DSTF functional.

\end{abstract}

\newpage
%\footnotetext[1]{E-Mail:
%\texttt{hainzl@doppler.thp.univie.ac.at}}

%\newpage

%\tableofcontents

\section{Introduction}

In this paper we study gradient correction terms for semiclassical
theories describing the ground state energies of heavy atoms in
strong homogeneous magnetic fields. Such an atom, with $N$
electrons of charge $-e$ and mass $m_e$ and nuclear charge $Ze$ is
described by the nonrelativistic Pauli Hamiltonian
\beq\label{PH}
H_{N}  =  \sum_{1 \leq j \leq N} \left\{ ((-i\nabla^{(j)} + {\bf
A}(x_{j}))\cdot {\bf \sigma}^{j})^{2} - \frac{Z}{|x_{j}|} \right\}
+  \sum_{1 \leq i<j \leq N} \frac{1}{|x_{i} - x_{j}|},
\eeq
acting on the Hilbertspace $\bigwedge_{1 \leq j \leq N} L^{2}({\R}^{3},
{\C}^{2})$ of electron wave functions. The units are chosen such that
$\hbar=2m_{e}=e=1$. The magnetic field is  ${\bf {B}} =
(0,0,B)$, with vector potential ${\bf {A}} =
\frac{1}{2}B(-x_2,x_1,0)$, where $B$ is the magnitude of the field
in units of $B_0 = \frac{m_{e}^{2}e^{3}c}{\hbar^{3}} = 2.35 \cdot
10^{9} $Gauss, the field strength for which the cyclotron radius
$l_{B} = (\hbar c/(eB))^{1/2}$ is equal to the Bohr radius $a_{0}
= \hbar^{2}/(m_{e}e^{2})$. The ground state energy is
\beq
\label{gse} E^{\rm Q}(N,Z,B) = \inf\{ (\psi,H_{N}\psi): \psi \in
{\rm domain} \,\, H_N, (\psi,\psi) = 1\}.
\eeq

In \cite{Liebetal1994S} Lieb, Solovej and Yngvason approximated
(\ref{gse}) by means of the MTF (Magnetic Thomas-Fermi) functional
\beq
\label {03} {\mathcal{E}} ^{\rm MTF}[\rho ] =   \int \tau_{B}(\rho
) -  \int V\rho + D(\rho ,\rho ),
\eeq
where $V(x) = Z/|x|$ and $D(\rho,\rho) =
\frac{1}{2}(\rho,|x|^{-1}\ast\rho)$. The magnetic energy density
$\tau_B$ is, by definition, the Legendre transform of the pressure
$P_B$, i.e.
\beq
\label {02} \tau_{B} (t) = \sup_{w \geq 0} [tw - P_{B}(w)],
\eeq
with
\begin{equation}\label{pres}
P_{B} (w) = \frac{B}{3 \pi ^{2}} (w^{3/2} - 2 \sum_{i=1}^{ \infty
} |2iB - w|_{-}^{3/2} ).
\end{equation}
The corresponding energy
\beq\label{mtf}
E^{\rm MTF}(N,Z,B) = \inf\{{\mathcal {E}}^{\rm MTF}[\rho] | \rho
\geq 0, \rho \in D^{\rm MTF}, \int\rho \leq N\}
\eeq
was proved by Lieb, Solovej and Yngvason to be asymptotically
exact, as shown in the following Theorem:
\begin{thm}
(\cite{Liebetal1994S} Theorem 5.1)
If $Z \to \infty$ with $N/Z$
fixed and $B/Z^{3} \rightarrow 0$, then
\begin{displaymath}
E^{\rm Q}(N,Z,B)/E^{\rm MTF}(N,Z,B) \rightarrow 1.
\end{displaymath}
\end{thm}
In the limit $B\to \infty$ the function $\tau_B$ is the kinetic
energy describing particles confined to the lowest Landau band,
i.e.,
\begin{displaymath}
\tau_{\infty}(t) = \frac{4\pi^{4}}{3} t^{3}/B^{2},
\end{displaymath}
which results in the STF (Strong Thomas-Fermi) functional
\beq
\label {04} {\mathcal{E}} ^{\rm STF}[\rho ] =
\frac{4\pi^{4}}{3B^{2}} \int \rho^{3} - \int V\rho + D(\rho ,\rho
).
\eeq
The corresponding energy $E^{\rm STF}$ is quantum mechanically
exact in the limit $Z\to \infty$ for $Z^{4/3} \ll B \ll Z^{3}$
(\cite{Liebetal1994S} Proposition 4.16), which emphasizes the
fact (\cite{Liebetal1994L} Theorem 1.2) that for $B \gg Z^{4/3}$ the
electrons are to leading order confined to the lowest Landau band.

\subsection{Corrections to the leading order of the full
Hamiltonian (\ref{PH})}

The best result to date concerning  corrections to the leading order of (\ref{gse})
is presented by Victor Ivrii in  \cite{Ivrii1996}, Theorem 0.2:
\begin{thm}(\cite{Ivrii1996} Theorem 0.2)\label{Ivrii0}
Let $ B \leq  Z^{3}$ and $N \sim Z$, then
\begin{equation}\label{Ivrii}
|E^{ Q}(N,Z,B) - E^{\rm MTF}(N,Z,B) - \frac{1}{4} Z^{2}| \leq R_{1} + R_{2},
\end{equation}
with
\begin{equation}\label{Ivrii2}
R_{1} = CZ^{4/3}(N + B)^{1/3} \quad and \quad R_{2} = CZ^{3/5}B^{4/5}.
\end{equation}
\end{thm}
\noindent
Recall the order of the energy, $E^Q \sim Z^{7/3}[1 +
B/Z^{4/3}]^{2/5}$.

We make a few comments concerning Ivrii's proof.
Let
\beq
H_{\bf A} = [(-i\nabla + {\bf
A}(x))\cdot {\bf \sigma}]^{2}
\eeq
denote the free Pauli-Hamiltonian and
\beq
\phi^{\rm MTF} = Z|\x|^{-1} - \rho^{\rm MTF} \ast |\x|^{-1}
\eeq
the self consistent magnetic TF potential.
The main part of the estimate (\ref{Ivrii}) is given by the
difference between
\beq\label{t1}
{\rm Tr}[H_{\bf A} - \phi^{\rm MTF} + \mu]_-,
\eeq
the sum of all negative eigenvalues of the operator $H_{\bf A} - \phi^{\rm MTF} +
\mu$, and its semiclassical approximation
\beq\label{t2}
\int P_B(\phi^{\rm MTF} - \mu).
\eeq
For those who are familiar with  microlocal analysis we
should remark that Ivrii does not really consider $\phi^{\rm MTF}$ but
a smooth mollification, which we also denote with
$\phi^{\rm MTF}$ for simplicity.
In order to derive accurate estimates of
\beq
\left|{\rm Tr}[H_{\bf A} - \phi^{\rm MTF} + \mu]_- - \int P_B(\phi^{\rm MTF} -
\mu)\right|
\eeq
Ivrii essentially divides the domain into two main
zones, for $B \geq Z^{4/3}$, namely
\beq
\chi_1 = \{\x| \ 0 \leq |\x| \leq B/Z \} \ \ {\rm and} \ \
\chi_2 = \{\x| \ B/Z \leq |\x| \leq
r_S = Z^{1/5}B^{-2/5} \}.
\eeq
A corresponding partition of unity is given by two function
$\varphi_1$ and $\varphi_2$, with $\varphi_1 +
\varphi_2 =1$ on $\chi_1 \cup \chi_2$ and $\varphi_i$
essentially supported in $\chi_i$.
Using scaling arguments and semiclassical spectral asymptotics
Ivrii treats each zone separately. In the inner zone $\chi_1$,
where all Landau levels are taken into account and the MTF
potential is very similar to the usual TF potential, he gets
\beq\label{x1}
\chi_1: \ \left|{\rm Tr}(\varphi_1 [H_{\bf A} - \phi^{\rm MTF} +
\mu]_-) - \int\varphi_1  P_B(\phi^{\rm MTF} - \mu) - \frac 14
Z^2\right| \leq R_1.
\eeq
We see that in $\chi_1$ the Scott correction is recovered.
Moreover, we should note that the machinery of semiclassical
spectral asymptotics can only be applied to $\chi_1$ under the
condition $Z/B \gg 1/Z$ which means that (\ref{x1}) is only valid
for $B \leq Z^{2-\delta}$, with arbitrary $\delta > 0$. For
$B\geq Z^2$ a semiclassical approximation is no longer possible
and the terms (\ref{t1}) and (\ref{t2}) have to be estimated
separately.
Since $R_2$ overcomes  $Z^2$ for $B \geq
Z^{7/4}$, we should point out that
the Scott correction in (\ref{Ivrii}) only provides the next to leading order for
$B \ll Z^{7/4}$, but in the domain $\chi_1$ it nevertheless makes
sense
up to $B\ll Z^2$ according to (\ref{x1}).

In the outer zone $\chi_2$ only the lowest Landau band is
occupied,
which implies that in this region the MTF energy is represented by
the STF energy corresponding to the functional (\ref{04}).
In $\chi_2$ Ivrii derives the estimate
\beq\label{x2}
\chi_2: \ \left|{\rm Tr}(\varphi_2 [H_{\bf A} - \phi^{\rm MTF} +
\mu]_-) - \int\varphi_2  P_B(\phi^{\rm MTF} - \mu)\right| \leq
R_2,
\eeq
where the main contribution of (\ref{x2}) really stems from the
edge of the STF atom $r_S \sim Z^{1/5}B^{-2/5}$.

For low magnetic fields ($B\leq Z$)
V. Ivrii even improves (\ref{Ivrii}) and recovers Dirac and
Schwinger corrections as well.
\begin{thm}(\cite{Ivrii1996} Theorem 0.3)\label{Ivrii3}
If $B\leq Z$ then
\beq
\left|E^{\rm QM}(N,Z,B) - E^{\rm MTF}(N,Z,B) - \frac{1}{4} Z^{2} +
c_{DS} \int (\rho^{\rm TF})^{4/3}\right| \leq
CN^{5/3}((1+B)/N)^\delta
\eeq
holds with some $\delta >0$ and an appropriate parameter $c_{DS}$.
\end{thm}

\subsubsection{The von Weizs\"acker term introduced to MTF}

The von Weizs\"acker correction term was successfully introduced
to the TF theory in the sense that it reproduces the Scott
correction (rigorously proven in
\cite{SiedentopWeikard1989P,Hughes1986}), i.e.
\beq
E^{\rm TFW} = E^{\rm TF} + O(Z^{2}) + o(Z^{2}).
\eeq

In addition to the $Z^2$ correction, the TFW theory remedies some
defects of the TF theory: The corresponding TFW density
is finite at the nuclei, binding of atoms occurs and negative ions
are stable, furthermore the density has exponential fall off at
infinity, at least for neutral atoms and molecules.

In the limit $B\to 0$ the function $\tau_B$ is the kinetic
energy density in zero magnetic field, i.e.
\begin{displaymath}
\tau_{0}(t) = \frac{3}{5}(3\pi^{2})^{2/3} t^{5/3}.
\end{displaymath}
Since for small values of $B$ the introduction of the
von Weizs\"acker term to the MTF functional is justified, we
will further check up to which values of $B$ this definition makes
sense. We get the functional
\beq
\label{MTFW} {\mathcal{E}}^{\rm MTFW}[\rho ] =  A\int | \nabla
\rho ^{\frac{1}{2}} |^{2} + \int \tau_{B} (\rho ) -  \int V\rho +
D(\rho ,\rho ),
\eeq
with a suitably chosen parameter $A$. The functional (\ref{MTFW}) can be treated
analogously to the usual TFW functional in
\cite{Benguriaetal1981}. So we will just sketch the proofs of the
main propositions.

It turns out that for $B \ll Z^{2}$ the
von Weizs\"acker term still produces the Scott correction, but
makes no longer sense for higher magnetic fields. We will derive
the following Theorem:
\begin{thm}\label{scott}
For all $B,Z$ and $N/Z$ fixed
\beq\label{scottg}
|E^{\rm MTFW} - E^{\rm MTF} - O(Z^{2})| \leq
CB^{4/5}Z^{3/5} + o(Z^2)
\eeq
\end{thm}
Remark: The estimate (\ref{scottg}) is clearly useful if $B\leq
Z^{7/4}$.
Theorem \ref{scott} will be proved in Section 2.2.

The Scott correction, just like  in TFW theory, comes from
distances of order $1/Z$ near the nucleus, whereas the bound
$CB^{4/5}Z^{3/5}$ comes from the edge of the MTF atom and dominates the
Scott correction for $B \geq Z^{7/4}$. Moreover,
(\ref{scottg}) is in accordance with Theorem \ref{Ivrii0}, which justifies a posteriori the
introduction of the von Weizs\"acker term to the MTF functional.

\subsection{Physics in the lowest Landau band}

The quantum mechanical ground state energy of particles confined
to the lowest Landau band is given by
\beq
\label{ce} E^{\rm Q}_{\rm conf}(N,Z,B) = \inf_{\parallel \psi
\parallel_{2} = 1}(\psi, \Pi^{N}_{0} H_{N}\Pi^{N}_{0} \psi).
\eeq
where $ \Pi_0$ represents the
projector on the lowest Landau band, given by the kernel
\beq
\label{pk} \Pi_{0}(x,x') =
\frac{B}{2\pi}\exp\left\{\frac{i}{2}(x_{\perp} \times x'_{\perp})
\cdot {\bf B} - \frac{1}{4}(x_{\perp} -
x'_{\perp})^{2}B\right\}\delta(x_{3} - x'_{3})P_\downarrow,
\eeq
where $P_\downarrow$ denotes the projection onto the spin down
component, and $\Pi^{N}_{0}$ denotes the $N$'th tensorial power of $\Pi_0$.
The leading order of $\Ec^Q$, as $B,Z \to \infty$ with $B \ll Z^3$, is given  by
$E^{\rm STF}$,
the ground state energy of the functional (\ref{04}).
In a companion work we show, what is expected by Ivrii's Theorem
\ref{Ivrii0},
\beq\label{stfq}
|\Ec^Q - E^{\rm STF}| \leq CB^{4/5}Z^{3/5}.
\eeq
As we have argued above the main contribution to the estimate
(\ref{stfq})
comes from the edge of the STF atom, $r_S \sim Z^{1/5}B^{-2/5}$.
Recall the order of $E^{\rm STF}$, $E^{\rm STF}[N,Z,B] =
Z^{3}(B/Z^{3})^{2/5}E^{\rm STF}[N/Z,1,1]$, so that the estimate is
only of interest for $B \ll Z^3$.

As a better approximation to $\Ec^Q$, valid also for $B\geq Z^3$,
Lieb, Solovej and Yngvason suggested a density matrix functional defined as
\begin{displaymath}
{\mathcal {E}}^{\rm DM}[\Gamma] = \int_{{\mathbb {R}}^{2}}{\rm
Tr}_{L^{2}({\mathbb {R}})}
[-\partial_{3}^{2}\Gamma_{x_{\perp}}]dx_{\perp} - Z\int
|x|^{-1}\rho_{\Gamma}(x) + D(\rho_{\Gamma},\rho_{\Gamma}).
\end{displaymath}
Its variable is an operator valued function
\begin{displaymath}
\Gamma: x_{\perp} \rightarrow \Gamma_{x_{\perp}},
\end{displaymath}
where $\Gamma_{x_\perp}$ is an integral operator on
$L^{2}(\mathbb{R})$, given by a kernel $\Gamma_{x_\perp}(x_3,y_3)$
and  satisfying
\begin{equation}
\label {opb} 0 \leq \Gamma_{x_{\perp}} \leq ({B}/{2\pi})I
\end{equation}
as an operator on $L^2({\mathbb {R}})$.
The energy
\begin{displaymath}
E^{\rm DM}(N,Z,B) = \inf\{{\mathcal {E}}^{\rm DM}[\Gamma]|
\,\Gamma \,\mbox{satisfies} \,(\ref {opb})\,\,{\rm and}\,\, \int
\rho_{\Gamma} \leq N\}
\end{displaymath}
turns out to be asymptotically exact for magnetic fields in the
following precise sense:
\begin{thm}(\cite{Liebetal1994L} Theorem 5.1 and 7.1 and equations 7.3 and
8.5)
For some constants, $C_\lambda$ and $C'_\lambda$, we have
\begin{equation}\label{dmqm}
R_U \geq E_{\rm conf}^{\rm Q}(N,Z,B)-E^{\rm DM}(N,Z,B) \geq -R_L,
\end{equation}
with
\begin{equation}\label{RL}
R_L = C_\lambda\min\{Z^{17/15}B^{2/5}, Z^{8/3}[1+(\ln(B/Z^{3}))^{2}]\}
\end{equation}
and
\[R_U=C'_\lambda\min\{Z^{5/3}B^{1/3},
Z^{8/3}[1+\ln(Z)+(\ln(B/Z^{3}))^{2}]^{5/6}\}.\]
\end{thm}
We remark that the STF energy is the natural semiclassical
approximation of the DM energy. More precisely, the DM
energy can be written as
\beq
\frac B{2\pi}\int d\xperp \Tr_{L^2(\R)}[-\partial^2_z - \phi_{\xperp}^{\rm DM} - \mu^{\rm DM}]_- +
\mu^{\rm DM} N - D(\rho^{\rm DM},\rho^{\rm DM}),
\eeq
whereas the STF energy is given by the corresponding semiclassical
expression
\beq\label{semex}
\frac B{2\pi}\int d\xperp \int \int dp dz [p^2 - \phi_{\xperp}^{\rm STF} -
\mu^{\rm STF}]_- + \mu^{\rm STF} N - D(\rho^{\rm STF},\rho^{\rm
STF}).
\eeq
With the decomposition
$L^2(\R^3,d\x;\mathbb{C}^2)=L^2(\R^2,d\xperp)\otimes L^2(\R,dz)\otimes\mathbb{C}^2$
the projector $\Pi_0$ can be written as
\beq\label{dec}
\Pi_0=\sum_{m\geq 0}|\phi_m\rangle\langle\phi_m|\otimes{1}\otimes P_\downarrow,
\eeq
where $\phi_m$ denotes the function in the
lowest Landau band with angular momentum $-m\leq 0$,
i.e., using polar
coordinates $(r,\varphi)$,
\beq\label{amf}
\phi_m(\xperp)=\sqrt\frac B{2\pi}\frac 1{\sqrt{m !}}\left(\frac
{Br^2}{2}\right)^{m/2}e^{-i m \varphi}e^{-B r^2/4}.
\eeq
Using this and $H_{\bf A} \Phi_m =0$, we can write
\beq
\Pi_{0}H_{\A}\Pi_{0} = \sum_{m\geq 0}|\phi_m\rangle\langle\phi_m |
\otimes (-\partial_z^2)\otimes P_\downarrow.
\eeq
Based on this decomposition the author and R. Seiringer introduced in \cite{HS1} a
natural modification of the DM functional
called {\it discrete density matrix functional}
(DDM)
\beq\label{ddmf}
\mathcal{E}^{\rm DDM}_{B,Z}[\Gamma]=\sum_{m\in\N_0}\left(\Tr[-\partial_z^2
\Gamma_m]-Z\int V_m(z)\rho_m(z)dz\right)+\widetilde D(\rho,\rho),
\eeq
where
\beq
\widetilde D(\rho,\rho)= \frac 12 \sum_{m, n}\int
V_{m,n}(z-z')\rho_m(z)\rho_n(z')\rmd z \rmd z',
\eeq
and the potentials $V_m$ and $V_{m,n}$ are given by
\beqa\nonumber
V_m(z)&=& \int \frac 1{|\x|}|\phi_m(\xperp)|^2 \rmd\xperp, \\ \label{pot}
V_{m,n}(z-z') &=& \int \frac {|\phi_m(\xperp)|^2 |\phi_n(\xperp
')|^2}{|\x-\x'|} \rmd\xperp \rmd\xperp'.
\eeqa
Here $\Gamma$ is a sequence of fermionic density matrices acting on
$L^2(\R,\rmd z)$,
\beq
\Gamma=(\Gamma_m)_{m\in\N_0},
\eeq
with corresponding densities $\rho=(\rho_m)_m$,
$\rho_m(z)=\Gamma_m(z,z)$. Note that $\ED$ depends on $B$ via the
potentials $V_m$ and $V_{m,n}$.
The corresponding energy is given by
\beq\label{ddme}
E^{\rm DDM}(N,Z,B)=\inf\left\{\ED[\Gamma]\left| \
\sum_m\Tr[\Gamma_m]\leq N\right.\right\}.
\eeq
It is shown in \cite{HS1} that $\Eddm$ correctly reproduces the confined ground state
energy $\Ec^{ Q}$ apart from errors due to the indirect part
of the Coulomb interaction energy:

\begin{thm}(\cite{HS1} Theorem 1.2)\label{11}
For some constant $c_\lambda$ depending only on $\lambda=N/Z$
\beq\label{rub}
0\geq  E_{\rm conf}^{\rm Q}(N,Z,B)-\Eddm(N,Z,B) \geq -R_L,
\eeq
with
\beq\label{rl}
R_L = c_\lambda\min\left\{Z^{17/15}B^{2/5},
Z^{8/3}(1+[\ln(B/Z^{3})]^{2})\right\}.
\eeq
\end{thm}

Since the functional (\ref{ddmf}) can also be seen as a reduced Hartree-Fock
functional, in the sense of
\cite{S91},
it does not surprise that the upper bound in (\ref{rub}) is an
improvement  to (\ref{dmqm}), the relation between
$E^{\rm DDM}$ and $\Ec^Q$.
In addition to better estimates, the DDM theory remedies the
defect of the DM theory having a sharply cut ground state density
supported in the set $\{\x|\ |\x| \leq \sqrt{2N/B}\}$, for the
respective three dimensional DDM density,
\beq
\rho^{\rm DDM}(\x) =
\sum_m \rho_m^{\rm DDM}(z)\phi^2_m(\xperp),
\eeq
has exponential
fall off at infinity. Furthermore, since the DDM energy describes
$\Ec^Q$ correctly apart from errors due to the  indirect part
of the Coulomb interaction energy, the DDM energy could give rise
to even recover the exchange term, by means of an improved lower
bound on the two body Coulomb repulsion for particles in the
lowest Landau band. For the exchange energy is anticipated to be
of order
$\ln(B/Z^{4/3})Z^{7/5}B^{1/5}$ for $B \ll Z^3$, which in
\cite{HS1} by
the author and R. Seiringer is conjectured to be
given by the term
\beq
c\ln(B/Z^{4/3})\sum_{i}  \int (\rho_i^{\rm DDM})^2.
\eeq
This would lead to the relation
\beq\label{ee}
\Ec^Q = E^{\rm DDM} - c\ln(B/Z^{4/3})\sum_{i}  \int (\rho_i^{\rm
DDM})^2+ o\left(\ln(B/Z^{4/3})Z^{7/5}B^{1/5}\right),
\eeq
with $c$ appropriately chosen.

We have stated above that the STF functional is the natural
semiclassical approximation of the DM functional. Hence, we can
ask for the natural semiclassical approximation of the
DDM functional, of which the answer is given by the
so called DSTF functional
\beq
\E[\rho]=\sum_{m\in\N_0}\left(\kappa\int \rho^{3}_m(z) - Z\int
V_m(z)\rho_m(z)dz\right)+\widetilde D(\rho,\rho),
\eeq
where $\rho$ is a sequence of one-dimensional densities,
$\rho=(\rho_m)_{m\in\N_0}$,
 $\kappa = \pi^2 /3$,
and the respective DSTF energy is defined as
\beq
E^{\rm DSTF}(N,Z,B)=\inf\left\{\E[\rho]\left| \ \sum_m\int \rho_m
\leq N\right.\right\}.
\eeq
In Section 3.2.2 we will argue that the DSTF functional is even the
natural semiclassical approximation of the ground state energy
$\Ec^Q$ itself.

\subsection{Gradient corrections for semiclassical lowest Landau band theories}

\subsubsection{The Tomishima-Shinjo correction term}

For higher magnetic fields, where only the
lowest Landau band has to be taken into account, Tomishima and
Shinjo \cite{Tomishima1979} obtained the gradient correction term
\beq
\label{22} \epsilon^{\rm TS}[\rho] = \frac{2\pi^{4}}{B^{3}}
\rho(\nabla_{\perp} \rho)^{2} - \frac{1}{3} (\nabla_{\parallel}
\rho^{1/2})^{2},
\eeq
i.e.
\beq \label{TS}
{\mathcal {E}}^{\rm TS}[\rho] =  \frac{2\pi^{4}}{B^{3}}\int
\rho(\nabla_{\perp} \rho)^{2} - \frac{1}{3}\int
(\nabla_{\parallel} \rho^{1/2})^{2} + {\mathcal {E}}^{\rm
STF}[\rho],
\eeq
by perturbation expansion of the canonical density matrix. In 1995
the authors of \cite{Mazzola1995} recovered the TS theory within
the framework of current density functional theory. Since (\ref{TS})
has a negative gradient correction along the magnetic field the
TS functional is no longer bounded from below. Hence the
corresponding energy cannot, as usual, be defined by minimizing
over a suitably domain of definition, but only through the solutions
of the corresponding Euler-Lagrange equation under the restriction
$\int \rho =N$, i.e
\beq \label{tse}
\frac{4\pi^{4}}{B^{2}}\rho^{2} -
\frac{\pi^{4}}{B^{3}}[(\nabla_{\perp} \rho)^{2} +
2\rho\Delta_{\perp}\rho] -
\frac{1}{12\rho}\left[\frac{1}{2\rho}(\nabla_{\parallel} \rho)^{2}
- \Delta_{\parallel}\rho\right]
 = V - \rho\ast \frac{1}{|x|}
- \mu(N).
\eeq
A direct attack on this complicated equation does not look
promising, but a rough estimate of the corrections to STF
can be obtained by inserting the density
 \beq
\rho(r) = \left\{\begin{array}{cc}
\rho^{\rm STF}(B^{-1/2})  &{\rm for} \quad r\leq
B^{-1/2},\\
\rho^{\rm STF}(r) &{\rm for} \quad r\geq
B^{-1/2}.
\end{array} \right.
\eeq
into (\ref{TS}).
The negative gradient term gives a correction
$-O(B^{4/5}Z^{3/5})$ coming from the edge of the STF atom.
From (\ref{stfq}) we know
that
\beq\label{QSTF}
|\Ec^Q - E^{\rm STF}| \leq CB^{4/5}Z^{3/5},
\eeq
where the main contribution also stems from $r_S \sim
Z^{1/5}B^{-2/5}$, the edge of the STF atom.

On the other hand
the positive gradient correction orthogonal to the magnetic field in
(\ref{TS})
produces a correction $O(B^{1/4}Z^{3/2})
$ at a distance
of order $B^{-1/2}$ from the nucleus.
This part of the correction can also be obtained from an
{\it isotropic Tomishima functional}, defined as
\begin{equation}
\label{C}
{\mathcal {E}}^{\rm IT}[\rho] = {\mathcal {E}}^{\rm STF}[\rho] +
\frac{2\pi^{4}}{B^{3}} \int \rho(\nabla \rho)^{2}.
\end{equation}
The functional (\ref{C}) has all the good
properties of the usual TF theories, such as convexity and boundedness from
below.
The study of this functional, which we do in detail in Section
\ref{k}, should help us to get a deeper understanding of the nature of
the positive correction term in (\ref{TS}).
For the ground state  energy of (\ref{C}) we will derive the following theorem.
\begin{thm}\label{Ct}
For all $B,Z$ and $N/Z$ fixed
\beq\label{cstf}
E^{\rm IT}(N,Z,B) - E^{\rm STF}(N,Z,B)  = O(B^{1/4}Z^{3/2}) +
o(B^{1/4}Z^{3/2}).
\eeq
\end{thm}
Furthermore
 we will argue in Section \ref{k} that the $\rho(\nabla
\rho)^{2}$ term remedies the defect of the STF theory that
the full
Coulomb potential is used although the particles in the lowest Landau
band   do not
see the full singularity, since they are smeared over a
region of radius $B^{-1/2}$. In contrast to TFW theory, where
the maximal number $N_c$ of electrons that can be bound is
strictly larger than $Z$, it will as a slight surprise turn
out that in IT theory  $N_c = Z$, just like in the STF
theory itself. Also, the radius of atoms in IT theory is
finite, as in STF theory. These features confirm that the $\rho(\nabla
\rho)^{2}$ term essentially only effects the  density
close to the nucleus.

\subsubsection{Gradient correction for the discrete STF theory}

As discussed above, the gradient term $\sim \rho(\nabla_\perp \rho)^2$
in (\ref{TS}) produces essentially a smearing of the Coulomb singularities
over a distance of the radius $B^{-1/2}$. The same effect was obtained by
replacing STF by DSTF.

It appears thus natural to look for a negative gradient term that
has an analogue effect in DSTF as the negative gradient term in
(\ref{TS}) has in STF theory, i.e. provides corrections at the
edge of the atom.

The DSTF theory is effectively a theory of coupled one-dimensional
problems. Analogous arguments as lead to the von Weizs\"acker term
for a three dimensional Fermi gas give for a one-dimensional Fermi
gas a gradient correction  $-\frac{1}{3} (\nabla \rho^{1/2})^{2}$
, cf. \cite{Shao1993}. Hence we suggest the definition of a {\it
discrete von Weizs\"acker} functional:
\beq
\label{DW} {\mathcal {E}}^{\rm DW}[\rho] = \sum_{m\in\N_0}\left(-
\frac{1}{3} \int |\partial_{z}\sqrt{\rho_m(z)}|^{2}+  \kappa
\int\rho_m^{3}(z) - Z\int V_m(z)\rho_m(z)\right) + \widetilde
D(\rho,\rho)
\eeq
By denoting $\sqrt{\rho_m}=\psi_m$ we arrive, under the restriction
$\sum_m\int \psi_m^{2}=N$, at the corresponding TF equation
\beq
\label {EMT} (1/3)\partial_{z}^{2}\psi_{n}(z) +
3\kappa\psi_{n}^{5}(z) = [\varphi_n(z) - \mu(N)]\psi_{n}(z)\quad
\forall n \in \N_0,
\eeq
with
\begin{displaymath}
\varphi_n(z) = ZV_n(z) - \sum_m \int\psi^{2}_m (z')V_{m,n}(z-z')dz'.
\end{displaymath}
These coupled equations are probably somewhat easier to deal with
than (\ref{tse}).

If  we reduce (\ref{DW}) to the angular momentum channel  $m=0$
and drop the Coulomb repulsion term,
we get the one-dimensional functional
\beq
\label{1DW} {\mathcal {E}}^{\rm 1DW}[\rho] = -
\frac{1}{3} \int |\partial_{z}\sqrt{\rho(z)}|^{2}+  \kappa
\int\rho^{3}(z) - Z\int V_0(z)\rho(z).
\eeq
This simplified  functional will be studied in Section 3.3.2.
In particular, we shall show that the negative gradient term
reproduces the right QM correction to the energy without the gradient term.

\section{The magnetic TFW theory}

\subsection{Mathematical analysis of the MTFW functional}

In this section we are going to mathematically analyze the MTFW
functional
\beq\label{2mtfw}
{\mathcal{E}}^{\rm MTFW}[\rho ] =  A\int | \nabla
\rho ^{\frac{1}{2}} |^{2} + \int \tau_{B} (\rho ) -  \int V\rho +
D(\rho ,\rho ).
\eeq
Since the mathematical propositions do not depend on the parameter
$A$,  we let $A$ be $1$ in this section. The most important
features of $\tau_B(t)$ which will be used in our calculations are
(compare \cite{Liebetal1994S} Lemma 4.1):
\begin{equation}
\label {taub} \tau_{B}'(t) \leq \kappa_{1}t^{2/3} \quad \mbox{and}
\quad \tau_{B}(t) \leq \frac{3}{5}\kappa_{1}t^{5/3},
\end{equation}
with $\kappa_1 = (4\pi^2)^{2/3}$.

Since the MTFW functional does not differ very much from the
functional in \cite{Benguriaetal1981}, where the authors used a
kinetic energy density $\tau(\rho)= (1/p)\rho^p$, our procedure in
analyzing (\ref{MTFW}) will be in analogy to their work. We are thus concerned
 with the minimizing
problem
\begin{equation}
\label{Minp} \mbox{Min} \{ {\mathcal{E}}^{\rm MTFW}[\rho ] | \rho
\geq 0, \rho \in L^{1} \cap L_{\rm loc}^{5/3}, \nabla \rho^
\frac{1}{2} \in L^{2} \,\,\mbox{and} \,\, \int \rho = N \},
\end{equation}
where $N$ is a positive constant, which physically is the total
charge number. Our main result is the following:
\begin{thm}
\label{mthm} There is a critical number $0 < N_c < \infty$, so
that
\begin{enumerate}
\item
if $N \leq N_c$ (\ref{Minp}) has a unique minimizer,
\item if $N > N_c$ (\ref{Minp}) has no minimizer,
\item $N_c > Z$.
\end{enumerate}
\end{thm}
Similar to \cite{Benguriaetal1981} we first examine the problem
\begin{equation}
\mbox{Min}\{ {\mathcal{E}}^{\rm MTFW}[\rho]| \rho \in D \}
\end{equation}
with
\begin{equation}\label{D}
D = \{ \rho | \rho \geq 0, \tau_{B}(\rho ) < \infty, \rho \in
L^{3}, \nabla \rho^{1/2} \in L^{2}, D(\rho ,\rho ) < \infty \}
\end{equation}
and proof the existence of a unique minimizer $\rho_0$. Since $D$
contains the domain of (\ref{Minp}) we have to
show $\rho_0 \in L^1(\mathbb{R}^3)$ in order to guarantee $N_c <
\infty$. Furthermore we will derive $N_c > Z$, which shows that this
theory allows negative ions. The  proofs
will be based on the Euler-Lagrange equation for
$\psi=\sqrt{\rho_0}$.

First we consider some basic properties of (\ref{2mtfw}).
\begin{lem}
For $D$ defined in (\ref{D}) we have
\begin{equation}
D \subset \{ \rho | \rho \geq 0, \rho \in L^{3} \cap L^{5/3}_{\rm
loc}, \nabla \rho^{1/2} \in L^{2}, D(\rho ,\rho ) < \infty \}
\equiv \bar D.
\end{equation}
\end{lem}
\begin{proof}
According to \cite{Liebetal1994S} (4.19) one gets for all $ \Omega
\subset {\R}^{3} $
\begin{displaymath}
\int_{\Omega} \rho(x)^{5/3}dx \leq
\frac{1}{\kappa_{3}}\int_{\Omega_{1}} \tau_{B}(\rho(x))dx  +
C\,{\rm Vol}(\Omega_{2}) < \infty
\end{displaymath}
with $ \Omega = \Omega_{1} \cup \Omega_{2}.  $
\end{proof}

\begin{prop}\label{ex}
The absolute minimum of  ${\mathcal{E}}^{\rm MTFW}[\rho] $ is  achieved for
a unique $\rho_{0} \in D$.
\end{prop}
\begin{proof}(cf. [BBL] Lemmas 2,3,4 and 5.)
Recall that, by definition (\ref{02}), $\tau_B(t)$ is strictly convex,
hence ${\mathcal{E}}^{\rm MTFW}[\rho]$ is strictly convex.

Let $\rho_{n} $ be a minimizing sequence. There exists a constant $C$ such that
\begin{displaymath}
\|\rho_{n} \|_{3} \leq C , \int \tau_{B}(\rho_{n}) \leq C ,
\|\nabla \rho_{n}^{1/2} \|_{2} \leq C , D(\rho_{n} ,\rho_{n} )
\leq C.
\end{displaymath}
By the Banach-Alaoglu theorem we can extract a subsequence, still
denoted as
$\rho_{n}$, with \beqa \label{1l3} &\rho_{n} \rightharpoonup
\rho_{0} \quad \mbox{weakly in} \quad L^{3},& \\ \label{1l2}
&\nabla\rho_{n}^{1/2} \rightharpoonup \nabla\rho_{0}^{1/2} \quad
\mbox{weakly in} \quad L^{2}.&
\eeqa
Since by use of H\"older's inequality  $\|\rho_{n}^{1/2}
\|_{H^{1}(\Omega)} \leq C(\Omega)$ and $H^{1}(\Omega )$ is
relatively compact in $L^{2}(\Omega)$, if $\Omega$ is a bounded
smooth domain, $\rho_{n}^{1/2}$ has a subsequence converging in
$L^2(\Omega)$. Using Cantor's diagonal trick on a sequence of
increasing $\Omega$'s we arrive at
\begin{equation}
\rho_{n}^{1/2} \rightarrow \rho_{0}^{1/2} \ \ \ \mbox{a. e.}. \ \
\
\end{equation}
By Fatou's Lemma we get
\begin{displaymath}
\liminf \int \tau_{B}(\rho_{n}) \geq \int \tau_{B}(\rho_{0})
\, \, \, {\rm and} \,\,\,
\liminf D(\rho_{n},\rho_{n}) \geq D(\rho_{0},\rho_{0}).
\end{displaymath}
Since $L^{p}$ norms are weakly lower semicontinuous,
\begin{displaymath}
\liminf \int |\nabla\rho_{n}^{1/2}|^{2} \geq \int
|\nabla\rho_{0}^{1/2}|^{2}.
\end{displaymath}
Moreover, one can show, in analogy to Proposition  \ref{exminc}, that
\begin{displaymath}
\int V\rho_{n} \rightarrow \int V\rho_{0},
\end{displaymath}
so we altogether
arrive at
\begin{equation}
\liminf {\mathcal{E}}^{\rm MTFW}[\rho_{n}] \geq {\mathcal{E}}^{\rm
MTFW}[\rho_{0}].
\end{equation}
The uniqueness follows from the strict convexity of ${\mathcal{E}}^{\rm MTFW}[\rho]$.
\end{proof}

For the minimizing $\rho_0$ we now can derive an Euler-Lagrange
equation. Denote $\psi= \sqrt{\rho_0}$.
\begin{prop}
The minimizing $\psi^{2} = \rho_{0}$ satisfies
\begin{equation}
\label{115} -\Delta \psi + \tau'_{B}(\psi^{2})\psi = \varphi \psi,
\end{equation}
in the sense of distributions, with $\varphi = V - \psi^{2} \ast
\frac{1}{|x|}$.
\end{prop}
\begin{proof} (cf. [BBL] Lemma 6.) Note that $\psi ^2 \in D$ implies
$\varphi\psi, \tau_B '(\psi^2)\psi \in L^1_{\rm loc}$, which gives
(\ref{115}) a meaning in the sense of distributions. Consider the
set
\begin{equation}
\tilde D \equiv \{\zeta | \zeta \in L^{6} \cap L^{10/3}_{\rm loc},
\nabla \zeta \in L^{2} \,\, \mbox{and} \,\,  D(\zeta^{2}
,\zeta^{2} ) < \infty \}.
\end{equation}
If $\zeta \in \tilde D$ then $\rho = \zeta^{2}   \in D $ and
\begin{displaymath}
{\mathcal{E}} ^{\rm MTFW}[\rho ] =  \int | \nabla \zeta |^{2} +
\int \tau_{B} (\zeta^{2} ) -  \int V\zeta^{2} + D(\zeta^{2}
,\zeta^{2}) \equiv \phi(\zeta).
\end{displaymath}
We find $\phi(\psi) \leq \phi(\zeta)$ for all $\zeta \in \tilde
D$. Let $\eta \in C^{\infty}_{0}$. Using the fact that
$\frac{d}{dt}\phi(\psi +t\eta)|_{t=0} = 0$, we easily arrive at
\begin{equation}
-\int \psi \Delta \eta  +  \int\tau'_{B}(\psi^{2})\psi \eta = \int
\varphi\psi \eta.
\end{equation}
\end{proof}
Starting from Equation (\ref{115}) we can now step by step gain
several properties for $\psi$.
\begin{lem}
$\psi $ is  continuous on  ${\mathbb{R}}^{3}$, more precisely $\psi
\in C^{0,\alpha}_{\rm loc}$ for all $\alpha \leq 1$.
\end{lem}
\begin{proof} (cf. [BBL] Lemma 7.)
Since (\ref{115}) yields $-\Delta \psi \leq \varphi\psi$,
with $\varphi\psi \in L^{2-\delta}_{\rm loc}$, one gets $\psi \in
L^{\infty}_{\rm loc}$. Again using (\ref{115}) the proposition follows
by means of standard elliptic regularity theory.
\end{proof}
\begin{prop}
$\psi \in L^{2}({\mathbb{R}}^{3})$
\end{prop}
\begin{proof}
(cf. \cite{Benguriaetal1981} Lemma 8.)
Assume, by contradiction, $\int \psi^{2} = \infty$. Then we can
choose an $r$, such that
\begin{displaymath}
\int_{|x| \leq r}\psi^{2}(x) \geq Z + 2\delta,
\end{displaymath}
for some $\delta > 0$. Therefore
\begin{displaymath}
\psi^{2}\ast|x|^{-1} \geq \int_{|x| \leq r} \psi^{2}(x)(|x| +
|y|)^{-1}dy \geq (Z + 2\delta)/(|x| + r),
\end{displaymath}
which gives us
\begin{displaymath}
\varphi(x) = V(x) - \psi^{2}\ast|x|^{-1} \leq \frac{Z}{|x| - r} -
\frac{Z + 2\delta}{|x| + r},
\end{displaymath}
with $|x| > r$. Thus there exists an $r_{1} > r$, such that for
$|x|
> r_{1}$
\begin{equation}\label{Absch}
\varphi(x) \leq - \delta |x|^{-1}.
\end{equation}
From (\ref{115}) we get
\begin{equation}\label{cpd1}
-\Delta \psi + \delta |x|^{-1}\psi \leq 0,
\end{equation}
for $|x| > r_{1}$. Now we choose a comparison density
\begin{displaymath}
\tilde \psi(x) = Me^{-2(\delta|x|)^{1/2}},
\end{displaymath}
which satisfies
\begin{equation}\label{cpd}
-\Delta\tilde \psi + \delta |x|^{-1}\tilde \psi \geq 0.
\end{equation}
Hence by (\ref{cpd1}) and (\ref{cpd})
\begin{displaymath}
-\Delta(\psi - \tilde \psi) + \delta |x|^{-1}(\psi - \tilde \psi)
\leq 0
\end{displaymath}
for $|x| \geq r_{1}$. We fix $M$ such that
\begin{displaymath}
\psi(r_{1}) \leq \tilde \psi(r_{1}).
\end{displaymath}
If $\psi \to 0$ for $|x| \to \infty$,
 we immediately get
\begin{equation}\label{Vergl}
\psi \leq \tilde \psi \quad \mbox{for} \quad |x| > r_{1}
\end{equation}
from the maximum principle. The fact that
$\int {\tilde \psi}^{2} < \infty$ and $\psi \in L^{\infty}_{\rm
loc}$  contradicts  our assumption.
Unfortunately, we only know that $\psi \to \infty$ as $|x| \to \infty$ in a
weak sense, namely $\psi \in L^6$, so the authors in [BBL] used a
variant of Stampaccia's method to verify the statement of the
Lemma, which also works in our case.
\end{proof}

Mimicking the proof of \cite{Benguriaetal1981} Lemma 10 and using
the fact that $\tau'_{B}(\psi^{2})\psi - \varphi \psi$ is
continuous but not differentiable we get
\begin{lem}
$\psi  >  0$ everywhere and $\psi \in C^{2}$, except at $x=0$.
\end{lem}
Using (\ref{taub})  in the proof of
\cite{Benguriaetal1981} Lemma 11 and afterwards following the proof of
Lemma 13 we additionally get
\begin{prop}
$N_c = \int
\psi^2 > Z$.
\end{prop}

Before concluding the proof of Theorem \ref{mthm}, we need a final
lemma, which is the equivalent to \cite{Benguriaetal1981} Lemma 14.
\begin{lem}\label{conc}
For every $N > 0$ we have  $ \inf\{ {\mathcal{E}}^{\rm MTFW}[\rho]
| \rho \in \bar D \, \mbox{and} \, \int \rho = N \} = \inf\{
{\mathcal{E}}^{\rm MTFW}[\rho] | \rho \in \bar D  \, \mbox{and} \,
\int \rho \leq N \}.$
\end{lem}
\noindent
{\em Proof of Theorem \ref{mthm}}:\\ For every $N$ we set
\begin{displaymath}
E(N) \equiv \inf\{ {\mathcal{E}}^{\rm MTFW}[\rho] | \rho \in \bar
D \, \mbox{and} \, \int \rho \leq N \}.
\end{displaymath}
Obviously $E(N)$ is non-increasing and convex. The same proof as in
Proposition \ref{ex} shows that there exists a $\rho_{N} \in \bar
D$ with $\int \rho_{N}  \leq N$ and
\begin{displaymath}
{\mathcal{E}}^{\rm MTFW}[\rho_{N}]  =  E(N).
\end{displaymath}
With $N_{c} = \int \psi^{2}$ it is clear that  $E(N)
$ is constant for $N > N_c$: $E(N) = E(N_{c})$, while $E(N)$ is strictly
decreasing on the interval $[0, N_{c} ]$. For $N \leq N_{c}$ :
$\int \rho_N = N$, which implies that (\ref{Minp}) has a unique
solution. On the other hand we deduce from Lemma \ref{conc} that
for $N
> N_{c}$ (\ref{Minp}) has no solution, which concludes the proof of
Theorem \ref{mthm}.   \hfill        $\qed$\\

After having guaranteed the existence of a minimizing density
$\rho_N$ for (\ref{Minp}), we can derive an Euler-Lagrange
equation under the variational  restriction $\int \rho = N$.
\begin{prop}

Denote $\psi = \rho_{N}^{1/2} $, with $\rho_{N}$ the minimizing
density for
\begin{displaymath}
\inf\{ {\mathcal{E}}^{\rm MTFW}[\rho] | \rho \in \bar D ,  \int
\rho = N \}
\end{displaymath}
under the restriction $N \leq N_{c}$. Then we have
\begin{equation}
\label{122} -\Delta \psi  +  \tau'_{B}(\psi^{2}) \psi  -  \varphi
\psi  =  \mu(N) \psi,
\end{equation}
where
\begin{equation}
\label{EN} \mu(N) = \frac{d}{dN} E(N).
\end{equation}
\end{prop}
\begin{proof}
The derivation of  (\ref{122}) works analogously to (\ref{115})
apart from the difference that $\mu$ is the Lagrange parameter for
the restriction
$\int\rho = N$.
We can infer
\begin{equation}
\label {CP} \left.\frac{d}{dt} {\mathcal{E}}^{\rm MTFW}[t \rho +
(1-t)\rho_{N}]\right|_{t=0} = \mu(N) \int (\rho_{N} -\rho),
\end{equation}
which implies by means of convexity of the functional
\begin{displaymath}
{\mathcal{E}}^{\rm MTFW}[\rho_{N}] - {\mathcal{E}}^{\rm
MTFW}[\rho] \geq \mu(N) \int (\rho_{N} -\rho),
\end{displaymath}
or equivalently for every $N'$
\begin{equation}
\label {au} E(N) - E(N') \geq \mu(N)(N - N').
\end{equation}
On the other hand we derive from (\ref {CP})
\begin{displaymath}
{\mathcal{E}}^{\rm MTFW}[t \rho + (1-t)\rho_{N}] -
{\mathcal{E}}^{\rm MTFW}[ \rho_{N}] = t\mu(N) \int (\rho_{N}
- \rho) + o(t),
\end{displaymath}
which yields with $\rho = 2\rho_{N}$ and $\rho = \frac{1}{2} \rho_{N}$, respectively,
\begin{equation}
\label {ao} E(N \pm tN) - E(N) \leq \pm \mu(N)tN + o(t).
\end{equation}
Hence (\ref{au}) and (\ref{ao})
together imply (\ref{EN}).
\end{proof}
Next we take a look at the behavior at infinity of the
minimizing densities $\rho_N$. At least for $N < N_c$ one gets
exponential decay.
\begin{prop}
(a) \,  Let $\mu < 0$, which is equivalent to $N < N_c$, then for every
$\delta > 0$, with $\mu < -\delta$, there exists a constant $M$,
such that, for the corresponding minimizer $\psi = \rho_N^{1/2}$,
\beq\label{exp}
\psi\leq M e^{-\delta^{1/2}|x|}.
\eeq
(b) \, Let $N = N_c$, then for every $\delta < N_c - Z$ there is
a constant $M$, such that
\beq
\psi\leq M e^{-2(\delta|x|)^{1/2}}.
\eeq
\end{prop}
\begin{proof}
Note, that we have not yet shown that $\psi \to 0$ as $|x| \to \infty$ in a
strong sense, for we only know $\psi \in L^2$.
From equation (\ref{122}) we derive
$-\Delta \psi \leq V\psi$, which implies $(-\Delta + I)\psi \leq (V +
I)\psi$. Since we know $(V + I)\psi \in L^2$, recall that
$(V + I) \in L^2 + L^\infty$ and $\psi \in L^2 \cap L^\infty$, we conclude
from
\beq
\psi \leq (-\Delta + I)^{-1}[(V+I)\psi],
\eeq
that $\psi \to 0$ at infinity, e.g. from the well known fact (\cite{LS} Lemma II.25)
that the convolution
$f\ast g$ of two functions
$f\in L^{p}, g\in L^{g}$, with $1/p + 1/q =1$, goes to $0$ at infinity.\\
(a) \,(cf. [L1] Theorem 7.24.) Let $\delta < -\mu$.
From (\ref{122}) we get
\beq
(-\Delta + \delta)\psi  = [- \tau'_{B}(\psi^{2})+ \varphi
+ \mu + \delta] \psi
\eeq
and
\beq
\psi  =(-\Delta + \delta)^{-1}[-  \tau'_{B}(\psi^{2})+  \varphi
+ \mu + \delta] \psi.
\eeq
Since $ V, \psi \to 0$ as $|x| \to \infty$, there is a $r_1$,  such
that $[-  \tau'_{B}(\psi^{2})+  \varphi
+ \mu + \delta] < 0$ for $|x| > r_1$.
This implies
\beq
\psi(x) \leq \int_{|y| \leq r_1} (4\pi|x - y|)^{-1}
e^{-\delta^{1/2}|x-y|}([- \tau'_{B}(\psi^{2})+  \varphi
+ \mu + \delta] \psi)(y) dy < \infty,
\eeq
and (\ref{exp}) with
\beq
M = \sup_x e^{\delta^{1/2}r_1}\int_{|y| \leq r_1} (4\pi|x - y|)^{-1}
([- \tau'_{B}(\psi^{2})+  \varphi
+ \mu + \delta] \psi)(y) dy,
\eeq
(b) \,\, This follows directly from (\ref{Vergl}) and the fact
that $\psi \to 0$ as $|x| \to \infty$.
\end{proof}

We state a final proposition concerning the behavior of the
chemical potential at $N=0$ (which is $-\infty$ in the usual TF
theory), because of the simple and illuminating proof.

\begin{prop}
Let $e_{0}= - \frac{1}{4} Z^{2}$ be the smallest eigenvalue of the
Schr\"o\-dinger operator $-\Delta
- Z/|x|$. Then
\begin{equation}
\label{1Abl} \left.\mu(0) = \frac{d}{dN}E(N)\right|_{N =0} = e_{0}.
\end{equation}
\end{prop}
\begin{proof}(cf. [BBL] Lemma 15.)
Let $\varphi(x)$ be the normalized eigenvector of $-\Delta
- Z/|x|$ belonging to the lowest eigenvalue $e_{0}= - \frac{1}{4}
Z^{2}$ and let $\rho_{N} = N\varphi(x)^{2}$. Then
\begin{eqnarray*}
E(N) \leq {\mathcal{E}}^{\rm MTFW}[\rho_{N}] &=&
\int|\nabla\rho_{N}^{1/2}|^{2} - \int V\rho_{N} + \int
\tau_{B}(\rho_{N}) + D(\rho_{N},\rho_{N})\\ &=&
N[(\varphi,-\Delta\varphi) - Z(\varphi,|x|^{-1}\varphi)] + \int
\tau_{B}(\rho_{N}) + D(\rho_{N},\rho_{N})\\ &\leq& N e_{0} +
C_{1}N^{5/3} + C_{2}N^{2}.
\end{eqnarray*}
On the other hand we have
\begin{eqnarray*}
E(N) &\geq& \inf_{\int\rho = N} \left\{\int|\nabla\rho^{1/2}|^{2}
- \int V\rho\right\}\\ &=& N\inf\mbox{spec}\{-\Delta - Z|x|^{-1}\}
= N e_{0},
\end{eqnarray*}
which altogether implies
\begin{displaymath}
\lim_{N \rightarrow +0}\frac{E(N)}{N } = e_{0}.
\end{displaymath}
Taking into account that $E(0) = 0$
this is equivalent to (\ref{1Abl}).
\end{proof}

\subsection{The Scott correction}

If one takes a look at Lieb's proof [L1] that in the usual TF theory without
magnetic fields the von Weizs\"acker term produces the Scott correction, one realizes
that
the main correction comes from distances of order $Z^{-1}$ from the nucleus.
It is thus reasonable to guess that in the MTF theory the von Weizs\"acker term
produces the  Scott correction as long as $\rho^{\rm MTF}$,
the density corresponding to the MTF energy (\ref{mtf}), is well approximated by the usual
TF density  $\rho^{\rm TF}$ up to distances of order $Z^{-1}$ from the nucleus. This
condition is equivalent to
the demand that $\tau_B'(\rho)$ is proportional to $\rho^{2/3}$ for $r\sim Z^{-1}$ and
this is the
case for $B \ll Z^2$.
In other words, for $B \ll Z^2$ the von Weizs\"acker term produces a $Z^2$ correction at
the distance
of order $Z^{-1}$ from the nucleus. At the edge of the MTF atom, the radius is  known
\cite{Liebetal1994S} to be proportional to $ Z^{1/5}B^{-2/5}$ and the lowest Landau
band is
occupied, which leads to $\tau_B'(\rho^{\rm MTF}) = 4\pi^4 B^{-2}(\rho^{\rm MTF})^2$
in the
outer region.
Hence, one computes very easily, by using $\rho^{\rm MTF}$ as
comparison density,
that the correction coming from the edge of the atom is of order
$B^{4/5}Z^{3/5}$.

Thus, for $B \geq Z^{7/4}$ the correction from the edge of the atom overcomes the Scott
correction
and Theorem \ref{scott} follows by using the variational density (\ref{varden}).
So it suffices to prove Theorem \ref{scott} for $B \leq Z^{7/4}$.\\

{\it Proof of Theorem \ref{scott}:}\\
First of all notice that there is an $r_B$, such that for $r \geq r_B$ the 
density
$\rho^{\rm MTF}$
corresponds to the lowest Landau band, i.e.
\beq
\label{trb}
\tau_B'(\rho^{\rm MTF}(r)) = \frac{4\pi^4}{B^2}(\rho^{\rm MTF}(r))^{2} \quad {\rm for}
\quad r\geq r_B,
\eeq
and for $r \leq r_B$
we have
\beq\label{tou}
\kappa_3 (\rho^{\rm MTF}(r))^{2/3} \leq
\tau_B'(\rho^{\rm MTF}(r)) \leq \kappa_1 (\rho^{\rm
MTF}(r))^{2/3},
\eeq
with $\kappa_3 = 0.83(3\pi^2)^{2/3}$.
Using (\ref{trb}) and (\ref{tou}) one realizes that if one fixes any $\eps  > 0$
with $B \leq Z^{2-\eps}$ there is a $\delta > 0$ such that
$r_B \geq Z^{\delta - 1}$. \\

{\em Lower bound}:\\
We know from Section 2.1 that there exists a $\rho_0$ satisfying
\beq
E^{\rm MTFW} = {\mathcal {E}}^{\rm MTFW}[ \rho_{0}] = \int \tau_B (\rho_{0}) +
\int |\nabla
\rho_{0}^{1/2}|^{2} - \int V \rho_{0} + D(\rho_{0},\rho_{0}).
\eeq
Denote
$Z|x|^{-1}= V = \tilde V + H $, with
\beqa\nonumber
H = Z/r - Z^{2}/b & \mbox{for} \quad r < b/Z
\quad \mbox{and} & 0 \, \, \,\mbox{otherwise},\\
\nonumber
 \tilde V = Z^{2}/b \qquad &
\mbox{for} \quad r < b/Z \quad \mbox{and} & Z/r \, \, \, \mbox{otherwise}.
\eeqa
Now let us rewrite the energy functional in the following way:
\begin{displaymath}
{\mathcal{E} }^{\rm MTFW}[ \rho_{0} ] = \int \tau_B ( \rho_{0} ) - \int \rho_{0}^{1/2}
\Delta \rho_{0}^{1/2} - \int \rho_{0}^{1/2} \tilde V
\rho_{0}^{1/2} - \int H \rho_{0} + D(\rho_{0},\rho_{0})
\end{displaymath}
Observe that
 $-\Delta - H$ $\geq \inf_\rho\{(\nabla \rho^{1/2},\nabla \rho^{1/2})  - (\rho^{1/2},H
 \rho^{1/2})\}$, which
by using Sobolev's inequality
can be bounded from below by
\begin{displaymath}
-\Delta - H \geq \inf_\rho\{
\parallel\rho\parallel_3^3 - \parallel \rho\parallel_3\parallel H\parallel_{3/2}\}.
\end{displaymath}
Since $\parallel H\parallel_{3/2} \sim b$
we can guarantee $-\Delta - H \geq 0$ with $b$ small enough.
Choosing such a $b$ we derive
\begin{eqnarray*}
E^{\rm MTFW} &\geq&  {\mathcal {E}}^{\rm MTF}[\rho_{0},\tilde V ] \\
&\geq & E^{\rm MTF}[\tilde V ] = \int \tau (\tilde \rho ) - \int \tilde \rho  \tilde V  +
D(\tilde \rho ,\tilde \rho )\\
&\geq & E^{\rm MTF}[V] + \int H
\tilde \rho.
\end{eqnarray*}
The density $\tilde \rho$ minimizes ${\mathcal {E}}^{\rm MTF}[ \rho,\tilde V ]$
and therefore fulfills the TF equation:
\begin{displaymath}
\tau_B '( \tilde \rho ) = \tilde V
- \tilde \rho \ast |x|^{-1}.
\end{displaymath}

By means of (\ref{tou}) we get that
$ \int H \tilde \rho =  O(Z^{2}) $ for $B \leq Z^{7/4}$
which yields
\begin{equation}\label{sl}
E^{\rm MTFW} \geq  E^{\rm MTF} + O(Z^{2}).
\end{equation}

{\em Upper bound}:\\
In order to get an upper bound we use a variational density $\rho$ in the following
way:
\beq\label{varden}
\rho(r) =
\left\{ \begin{array}{cc}
\bar \rho^{\rm TF}(Z^{-1}) & {\rm for} \quad r < Z^{-1},\\
\bar \rho^{\rm TF}(r) & {\rm for} \quad Z^{-1} \leq r \leq r_B,\\
  \rho^{\rm MTF}(r) & {\rm for} \quad  r \geq r_B,
  \end{array} \right.
\eeq
where  $\bar \rho^{\rm TF}$  indicates that  we essentially use the
usual TF density $\rho^{\rm TF}$. Only in the region $\eps Z/B \leq r \leq r_B$
we eventually have to modify $\rho^{\rm TF}$, such that
\begin{displaymath}
|E^{\rm MTF} - {\mathcal{E}}^{\rm MTF}[\rho]| \leq O(Z^{2}).
\end{displaymath}
(This e.g. can be done by following the way of \cite{Ivrii1996}
and taking the $C^\infty$ modification $W$ of the effective
potential $\phi^{\rm MTF}$ and defining $4\pi\bar \rho^{\rm TF}= \Delta(W - Z
|x|^{-1})$.)
By means of our variational density $\rho$  we get
\beqa
&\int_{r\leq r_B} |\nabla \rho^{1/2}|^2 = O(Z^2),&\\
&\int_{ r \geq r_B } |\nabla \rho^{1/2}|^2 = O(B^{4/5}Z^{3/5}).&
\eeqa
For $B \leq Z^{7/4}$ this leads to
\begin{displaymath}
E^{\rm MTFW} \leq E^{\rm MTF} + O(Z^{2}),
\end{displaymath}
which together with (\ref{sl}) completes the proof of Theorem \ref{scott}.
\hfill $\qed$

\section{Gradient corrections for STF type theories}

\subsection{The functional (\ref{C})}
\label{k}
Starting point in this section is the  functional
\begin{equation}
\label{Cf}
{\mathcal {E}}^{\rm IT}[\rho] = \frac{1}{B^{2}} \int \rho^{3} +
\frac{1}{B^{3}}\int
(\nabla \rho^{3/2})^{2} - \int V\rho + D(\rho,\rho),
\end{equation}
with $V$ and $D(\rho,\rho)$ defined as in (\ref{03}).
Compared to (\ref{C}) we rewrite
\begin{displaymath}
\frac{1}{B^{3}} \int \rho(\nabla \rho)^{2} =
\frac{4}{9B^{3}}\int
(\nabla \rho^{3/2})^{2},
\end{displaymath}
and for simplicity forget about the numerical constant, which does not  effect any
mathematical statements.
The corresponding energy is defined as
\begin{equation}
\label{29}
E(N,Z,B) = \inf\{ {\mathcal {E}}^{\rm IT}[\rho]| \rho \in \tilde D,
 \int \rho \leq N \},
\end{equation}
where the domain $\tilde D$ is given by
\begin{displaymath}
\tilde D = \{ \rho | \rho \geq 0, \rho \in L^{1} \cap
L^{3}, \nabla \rho^{3/2} \in L^{2} \}.
\end{displaymath}
In analogy to Section 2 we first consider the problem
\beq
{\rm Min} \{ {\mathcal{E}}^{\rm IT}[\rho]|\, \rho \in  D\},
\eeq
with
\begin{equation}
D = \{ \rho | \rho \geq 0, \rho \in L^{3}, \nabla \rho^{3/2} \in L^{2}, D(\rho,\rho)
< \infty \}
\end{equation}
and show that the minimum is achieved for a unique $\rho_0$.
By means of the corresponding Euler-Lagrange equation we shall deduce that $\rho_0$
is in $L^1(\R^3)$, more precisely $\int \rho_0 = Z$.

First of all, we collect some properties of (\ref{Cf}).
\begin{lem}\label{33}
There are positive constants $\alpha,\, C$, so that
\begin{equation}
{\mathcal{E}}^{\rm IT}[\rho] \geq \alpha (\| \rho \|_{3}  + \int
\tau_{B}(\rho ) + \|\nabla \rho^{1/2} \|^{2}_{2}  +  D(\rho ,\rho
) )  -  C,
\end{equation}
\end{lem}
\begin{proof} This is a consequence of
Lemma 2 in \cite{Benguriaetal1981}, which tells us that for every
$\varepsilon > 0$ there exists a constant $C_\eps$ so that
\begin{displaymath}
\int V\rho \leq \varepsilon\parallel \rho\parallel_{3} +
C_{\varepsilon}D(\rho,\rho)^{1/2}
\end{displaymath}
for every  $\rho \geq 0$.
\end{proof}
\begin{lem}
${\mathcal{E}}^{\rm IT}[\rho]$ is strictly convex in $\rho$.
\end{lem}
\begin{proof}
This follows immediately from the strict convexity of $(\nabla \rho^{3/2})^{2}$
and ${\mathcal{E}}^{\rm STF}$.
\end{proof}
By means of these Lemmas we can prove the existence of a unique
minimizer in $\tilde D$.
\begin{prop}\label{exminc}
The Minimum of ${\mathcal {E}}^{\rm IT}[\rho]$ is achieved by a unique
$\rho_{0} \in \tilde D$.
\end{prop}
\begin{proof} The proof is similar to that of Proposition 2.3.
Let $\rho_{n}$ be a minimizing sequence. By Lemma \ref{33}
we have
\begin{equation}
\Vert \rho_{n} \Vert_{3}^{3} \leq C,\quad \Vert \nabla \rho_{n}^{3/2}
\Vert_{2}^{2} \leq
C, \quad D(\rho_{n},\rho_{n}) \leq C.
\end{equation}

With Banach-Alaoglu theorem we therefore extract a subsequence, still
denoted by  $\rho_{n}$, such that
\begin{eqnarray}\label{cl3}
& \rho_{n} \rightharpoonup \rho_{0} \, \qquad
\mbox{weakly} \, \quad \mbox{in} \, \quad
L^{3},&\\
&\nabla \rho_{n}^{3/2} \rightharpoonup \nabla \rho_{0}^{3/2} \, \qquad
\mbox{weakly} \,\quad
\mbox{in} \, \quad L^{2}.&
\end{eqnarray}
Furthermore $\rho_{n}^{3/2}$  is bounded in $H^{1}$, which implies that
there exists a further subsequence, again denoted as $\rho_{n}$, with
\begin{displaymath}
\rho_{n}^{3/2} \rightarrow \rho_{0}^{3/2} \quad \mbox{a. e.}.
\end{displaymath}
(This relies on the fact that for a smooth bounded domain $\Omega$, $H^1(\Omega)$
is relatively compact in $L^2(\Omega)$.)
Hence, using Fatou's Lemma we get
\begin{displaymath}
\lim \inf D(\rho_{n}, \rho_{n}) \geq D(\rho_{0}, \rho_{0}),
\end{displaymath}
and by the weak lower semicontinuity of $L^{p}$-norms  we deduce
\begin{eqnarray*}
&\lim \inf \int (\nabla \rho_{n}^{3/2})^{2} \geq
\int (\nabla \rho_{0}^{3/2})^{2},&\\
&\lim \inf \int \rho_{n}^{3} \geq \int \rho_{0}^{3}.&
\end{eqnarray*}

In order to prove $\int V \rho_{n} \rightarrow \int V \rho_{0}$,
we decompose $V = V_{1} + V_{2}$ such that both functions are in $C^{\infty}$.
With $ V_{1} \in L^{3/2}$, (\ref{cl3}) implies
\begin{displaymath}
\int V_{1} \rho_{n} \rightarrow \int V_{1} \rho_{0}.
\end{displaymath}

On the other hand $V_{2}$ fulfills
\begin{displaymath}
\int V_{2} \rho_{n} = \int V_{2}[- \Delta(\rho_{n} \ast |x|^{-1})]
= \int (- \Delta V_{2})(\rho_{n} \ast |x|^{-1}),
\end{displaymath}
which converges to
\begin{displaymath}
\int (-\Delta V_{2})(\rho_{0} \ast
|x|^{-1}) = \int V_{2} \rho_{0},
\end{displaymath}
for $-\Delta V_{2} \in L^{6/5}$ and $\parallel \rho_{n} \ast |x|^{-1}
\parallel_{6}$ is bounded.

Thus
\begin{displaymath}
\lim \inf {\mathcal {E}}^{\rm IT}[\rho_{n}] \geq
{\mathcal {E}}^{\rm IT}[\rho_{0}].
\end{displaymath}
The uniqueness is an immediate consequence of the strict convexity of the
functional.
\end{proof}

For this minimizing $\rho_0$ we can derive an Euler-Lagrange equation.

\begin{prop}
The minimizing $\psi^{2/3} = \rho_{0}^{3/2}$ satisfies
\begin{equation}
\label{214}
(-\Delta + B)\psi = \frac{ B^{3}}{3} \varphi \psi^{-1/3},
\end{equation}
with $\varphi = V - \psi^{3/2} \ast \frac{1}{|x|}$, in the sense of distributions,
on the set where $\psi > 0$.
\end{prop}
\begin{proof} The uniqueness of the minimum, the spherical symmetry of the
functional (\ref{Cf}) and the fact that $\psi \in H^{1}$ implies
the continuity of $\psi$ away from the origin.
With $\varphi \in
L^{2}_{\rm loc}$ the Equation (\ref{214}) has a meaning in the
sense of distributions on the domain $\{x|\psi(x) > 0\}$.
Consider the set
\begin{displaymath}
\bar D = \{\eta|\eta \in H^{1}, D(\eta^{2/3}, \eta^{2/3}) < \infty\}.
\end{displaymath}
If $\eta \in \bar D$, then $\rho = (\eta^{2})^{1/3} \in D$
and
\begin{displaymath}
{\mathcal {E}}^{\rm IT}[\rho] = \frac{1}{B^{2}} \int \eta^{2} + \frac{1}{B^{3}}
\int(\nabla \eta)^{2} - \int V \eta^{2/3} + D(\eta^{2/3}, \eta^{2/3})
\equiv \phi(\eta).
\end{displaymath}
For all $\eta \in \bar D$ we find $\phi(\psi) \leq \phi(\eta)$.

Let $\xi \in C_{0}^{\infty}$, then using $\frac{d}{dt}\phi(\psi + t\xi)|_{t=0} = 0$
we infer
\begin{displaymath}
-\int \psi \Delta \xi + B\int \psi \xi = \frac{B^{3}}{3} \int \varphi
\psi^{-1/3} \xi.
\end{displaymath}
\end{proof}
\begin{prop}
$ \psi$ is bounded and  $\psi$ is in $C^{\infty}$ away from the
origin and eventual points with $\psi(x) = 0$.
\end{prop}
\begin{proof}
Denote
\begin{displaymath}
\Omega_{\epsilon} = \{x|\psi(x) \geq \epsilon \}.
\end{displaymath}
On this domain we have
\begin{displaymath}
 (-\Delta + B)\psi = f,
\end{displaymath}
with $f\in L^2_{\rm loc}$,
since $\varphi \in L^{2}_{\rm loc}$. So
we conclude from standard elliptic arguments (e.g. [LL] Section 10)
that $\psi$ is
bounded, hence continuous everywhere.
From
\begin{displaymath}
\Delta \varphi = 4\pi(\psi^{2/3}(x) - Z\delta(x))
\end{displaymath}
we get the two times differentiability of $\varphi$
away from the origin and as long as $\psi > 0$. By means of (\ref{214})
and a standard bootstrap argument we conclude
$\psi \in C^{\infty}$.
\end{proof}

\begin{prop}\label{neu}
$\int \psi^{2/3} = Z$.
\end{prop}
\begin{proof}
Suppose by contradiction
$\int \psi^{2/3} = \lambda
\not= Z$. We do not assume  $\lambda$ to be finite.
Then, as in  (\ref{Absch}), we get that there is some $r_1$
and an $\epsilon\ > 0$,
such that
\begin{equation}\label{e}
|\varphi(x)| \geq \epsilon/r,
\end{equation}
for $|x| \geq r_1$.
Since $\psi \in L^2$, we conclude $\psi < M/r^{3/2}$ for
$r$ large enough.
Hence, from
(\ref{214}) we get
\begin{displaymath}
|(-\Delta + B)\psi| = |\varphi \psi^{-1/3} | \geq \epsilon r^{-1/2},
\end{displaymath}
which is a contradiction to $\psi \in H^1$.
\end{proof}

\begin{rem}
The Proposition \ref{neu} is interesting, since gradient
corrections usually give rise to binding of additional electrons.
The reason that this is not the case in \ref{neu} relies on the
negative potent of $\psi$ on the right side of
(\ref{214}). This fact forces the potential $\varphi$ to
fall off much faster than $O(1/r)$.
\end{rem}
Since we only consider atoms with a point nucleus, we get the following
remark for
the minimizing $\rho_0$.
\begin{prop}
$\rho_0$ is a symmetric non increasing function of $|\x|$.
\end{prop}
\begin{proof}
As we have already argued above, the symmetry of $\rho_0$ follows from the
symmetry of the functional and the
uniqueness of $\rho_0$.

Denote $\rho_0^{\ast}$ the non increasing rearrangement of
$\rho_0$. (For definition see e.g. \cite{LL} Section 3.3.)
From the fact that $\int\rho_0 \leq Z$ and [L1]
Theorem 2.12 we get $\mathcal{E}^{\rm
STF}[\rho_0^{\ast}] \leq \mathcal{E}^{\rm STF}[\rho_0]$.
\cite{LL} Lemma 7.17 implies
\beq
\int |\nabla (\rho_0^{3/2})^{\ast} |^2\geq
\int |\nabla (\rho_0^{3/2})|^2
\eeq
and again from \cite{LL} 3.3 (v) we get
$ (\rho_0^{3/2})^{\ast} = (\rho_0^{\ast})^{3/2}$, which proves the
statement.
\end{proof}
\begin{prop}
$\psi $ has compact support.
\end{prop}
\begin{proof} Inserting $\Delta \varphi = \psi^{2/3}$ into (\ref{214}) yields the following equation for the
potential, away from the origin:
\begin{equation}
\label{215}
(\Delta \varphi)^{1/2}[\Delta(\Delta \varphi)^{3/2}] =
(\Delta \varphi)^{2} - \varphi,
\end{equation}
where we replaced the constants by one.
Since  $\varphi$ is spherical symmetric we can use the ansatz $\varphi =
\frac{1}{r} \chi(r)$ and obtain by (\ref{215}) the following fourth order
equation:

\begin{eqnarray}\nonumber
&2[\frac{1}{r^{4}} \chi''^{2} - \frac{2}{r^{3}} \chi''\chi'''
+ \frac{1}{r^{2}} \chi'''^{2}]^{2} +
[\frac{1}{r} \chi'']^{2}[\frac{1}{r} \chi'''']^{2}&\\
\nonumber
& + [\frac{1}{r^{4}} \chi''^{2} - \frac{2}{r^{3}}\chi''\chi''' +
\frac{1}{r^{2}} \chi'''^{2}][\frac{1}{r} \chi''][\frac{1}{r} \chi'''']&\\
\label{chi}
&= \frac{4}{9}[\frac{1}{r^{2}} \chi^{2} - \frac{2}{r^{3}}\chi\chi''^{2}
+ (\frac{1}{r} \chi'')^{4}]&
\end{eqnarray}

We can see that in the surrounding of each point $r_{0} > 0$ there exists a local
solution
\begin{displaymath}
\chi = a_{6}(r - r_{0})^{6} + O((r - r_{0})^{7}),
\end{displaymath}
with $a_{6} = ({\rm const}.)r_{0}^{2}$.
Since $\chi \equiv 0$ is also a solution of (\ref{chi}), every composed function
\begin{displaymath}
\chi =  a_{6}(r - r_{0})^{6} + O((r - r_{0})^{7}) \quad \mbox{for} \quad
r \leq r_{0} \quad \mbox{and} \quad \chi \equiv 0 \quad \mbox{for} \quad 
r > r_0
\end{displaymath}
is a local solution around $r_{0}$.

Away from $0$ the solutions can be uniquely continued up to $r =
0$.
Hence, there exists a solution $\chi$ and a $r_{1} > 0$,
such that $\chi(0) = Z$ and ${\rm supp}\chi = [0,r_{1}] $.

Repeating the argument of [L1] Theorem 2.6, one can show
that a solution of (\ref{214}), with $\psi \in H^{1}$ and $\int \psi^{2/3} < \infty$,
$(\psi^{2})^{1/3}$ uniquely determines the minimum of the functional (\ref{Cf}).
Therefore $\chi(r)$ uniquely determines the selfconsistent
potential $\phi = \chi/r$, which implies that
$\psi = \rho_{0}^{3/2} = (\Delta\phi)^{3/2}$ has compact support,
too.
\end{proof}
\begin{rem}
The preceding three propositions are equivalent to those for the
STF theory. This confirms that the $\rho(\nabla \rho)^{2}$ term
only amounts to changes close to the nucleus.
\end{rem}
By the convexity of the functional (\ref{Cf}) one easily derives the
following properties for the energy $E^{\rm IT}(N,Z,B)$:

\begin{prop}
$E^{\rm IT}(N,Z,B)$ is convex as a function of $N$  and strictly
monoton decreasing on the interval $[0,Z]$. For $N > Z$ we get $E^{\rm
IT}(N,Z,B)= E^{\rm IT}(Z,Z,B)$.
\end{prop}

Next we prove Theorem \ref{Ct}.\\
{\it Proof of Theorem \ref{Ct}:} {\em Upper bound:}\\
We use the comparison density  $\bar \rho$, with
\beq
\bar \rho(r) = \left \{\begin{array}{cc} \rho^{\rm STF}(l_{B})
& {\rm for} \,\,
r \leq l_{B}= B^{-1/2}\\ \rho^{\rm STF}(r)& {\rm otherwise}
\end{array} \right.
\eeq
and
immediately get
\begin{displaymath}
E^{\rm IT} \leq
E^{\rm STF} + O(Z^{3/2}B^{1/4}).
\end{displaymath}

{\em Lower bound}:\\
Let $\rho$ be the minimizer of the energy (\ref{29}), for given
$B,Z,N$.
We can rewrite
\begin{eqnarray*}
E^{\rm IT} &=& {\mathcal {E}}^{\rm IT}[\rho] \\
&=& \frac{\kappa}{B^{2}} \int \rho^{3} +
\frac{1}{B^{3}}\int|\nabla \rho^{3/2}|^2 - \int \tilde H \rho -
\int \tilde V \rho + D(\rho,\rho),
\end{eqnarray*}
where $\tilde H = \frac{Z}{r} - \frac{B^{1/2}Z}{b}$ for $r \leq
bl_{B}$ and $0$ otherwise,
$\tilde V = \frac{B^{1/2}Z}{b}$ for $r \leq bl_{B}$ and $Z/r$
otherwise.
Looking at the term
\begin{displaymath}
\frac{1}{B^{3}}\int|\nabla \rho^{3/2}|^2 - \int \tilde H \rho,
\end{displaymath}
we infer by means of the Sobolev inequality the estimate
\begin{equation}
\label{24}
\frac{1}{B^{3}}\int|\nabla \rho^{3/2}|^2 - \int \tilde H \rho \geq
\frac{4}{9B^{3}} \parallel \rho\parallel_{9}^{3} - \parallel\tilde H\parallel_{9/8}
\parallel\rho\parallel_{9}.
\end{equation}
Since $\parallel\tilde H\parallel_{9/8} \sim b^{15/9}$ we can
choose $b$ such that
$(\ref{24}) > 0$. Hence
\begin{eqnarray*}
E^{\rm IT} &\geq &
\frac{\kappa}{B^{2}} \int \rho^{3}  -
\int \tilde V \rho + D(\rho,\rho)\\
&\geq& E^{\rm STF}[\tilde V] = {\mathcal {E}}^{\rm STF}[\tilde V,\tilde \rho] \\
&\geq& E^{\rm STF} + \int \tilde \rho \tilde H.
\end{eqnarray*}
Since $\int \tilde \rho \tilde H =  O(Z^{3/2}B^{1/4})$
we prove the proposition.
\hfill $\qed$
\\

We see that the main contribution of the correction (\ref{C})
comes from the radius $l_B = B^{-1/2}$, in other words the gradient
correction repairs the infinity of the STF density at a distance
$B^{-1/2}$ from the nucleus.
This infinity stems from the fact that in the STF theory the full
$|x|^{-1}$ potential is involved, although particles in the lowest Landau band,
which are smeared over a radius of at least $B^{-1/2}$,  never
see the full Coulomb singularity.

Moreover, the Tomishima-Shinjo correction
(\ref{TS})
orthogonal to the magnetic field remedies the singularity
of the Coulomb potential in a similar way as the isotropic
gradient term.  We will see that the  same
effect, as caused
by (\ref{Cf}) and (\ref{TS}),
is also naturally
obtained by using the DSTF functional.

\subsection{A discrete von Weizs\"acker functional}

\subsubsection{The DSTF functional}

First of all we are going to collect some information about the
DSTF functional (which are rigorously proved in a companion
work). The DSTF functional
\beq\label{dstf}
\E[\rho]=\sum_{m\in\N_0}\left(\kappa\int \rho^{3}_m(z) - \int
V_m(z)\rho_m(z)dz\right)+\widetilde D(\rho,\rho),
\eeq
with $V_n$ and $V_{m,n}$ as in (\ref{pot}),
is defined on the domain
\beq
D= \{ \rho | \ \sum_m \int\rho^{3}_m < \infty,\sum_m \int \rho_m
< \infty, \widetilde D(\rho,\rho) <\infty\}
\eeq
with corresponding energy
\beq\label{edstf}
E^{\rm DSTF}(N,Z,B)=\inf\left\{\E[\rho]\left| \ \rho \in D \ \mbox{and} \ \
\sum_m\int \rho_m
\leq N\right.\right\}.
\eeq
Following the considerations of \cite{Liebetal1994S} one can easily see
that (\ref{dstf}) is convex and bounded from below on $D^{N} =
\{\rho|\ \rho \in D, \ \sum_m\int \rho_m \leq N\}$ and derive the
following Theorem:
\begin{thm}
With $N\leq Z$ fixed there exists a unique minimizer $\rho^{N}$ for
$\E$, under the restriction $\sum_m \int \rho_m \leq N$, i.e. $\Ed
(N,Z,B) = \E[\rho^{N}]$. Moreover, $\rho^{N}$ satisfies $\sum_m \int
\rho_m^{N} = N$.
\end{thm}
Furthermore each minimizer $\rho^{N}$ obeys the coupled TF equations
\beq
3\kappa (\rho_m^{N}(z))^{2}= [ZV_m(z) - \sum_n \int
V_{m,n}(z-z')\rho_n^{N}(z') + \mu(N)]_+ \ \ \forall (m\in \N_0),
\eeq
where $\mu(N)$ is the Lagrange parameter belonging to the
restriction $\sum_m \int
\rho_m = N$ and $[]_+$, with $[t]_+=t$ for $t\geq 0$ and $[t]_+
=0$ otherwise, corresponds to the fact that the functional is only varied
over positive functions, i.e. $\rho_n \geq 0$ $\forall n\in\N_0$.

By means of the notation
\beq
ZV_m(z) - \sum_n \int
V_{m,n}(z-z')\rho_n^{N}(z') + \mu(N) = \varphi^{(m)}_{\rm eff}(z)
\eeq
and inserting in the TF equation
we can rewrite the energy $\Ed
(N,Z,B) = \E[\rho^{N}]$ as
\beq\label{Phd}
\Ed(N,Z,B)= \sum_m\int\int [p^{2}-\vph]_-\frac{dpdz}{2\pi} + N\mu
-\widetilde
D(\rho^{N},\rho^{N}).
\eeq

\subsubsection{Semiclassical approximation of $\Ec^{Q}$}

First of all we want to state a useful theorem concerning the sum
of the negative eigenvalues of the one-particle operator $\Pi_0
[H_A + \varphi]\Pi_0$, where $H_A =  ((-i\nabla + {\bf A}(x))\cdot
{\bf \sigma})^{2}$ and $\varphi$ is an axialy symmetric potential
$\varphi(r,z)$ with $r=|\xperp|$.

\begin{thm}\label{tre}
Let $\varphi = \varphi(r,z)$ be axially symmetric. Then one can
write the trace of the negative part of the operator $\Pi_0 [H_A +
\varphi]\Pi_0$ as the sum of one-dimensional traces, i.e.
\beq
\Tr[\Pi_0
[H_A + \varphi]\Pi_0]_- =
\sum_{m\in\N_0}\Tr_{L^2(\mathbb{R})}[-\partial^2_z + \varphi^{(m)}(z)]_-,
\eeq
with
\beq
\varphi^{(m)}(z) = \int \varphi(x)|\phi_m(\xperp)|^2 d\xperp.
\eeq
\end{thm}
\begin{proof}
Let $L_z$ denote the angular momentum operator parallel to the
magnetic field. Since $\varphi= \varphi(r,z)$, we have
\beq
\left [\Pi_0
[H_A + \varphi]\Pi_0, L_z\right] = 0,
\eeq
which implies that the eigenvectors of the operator
$\Pi_0
[H_A + \varphi]\Pi_0$ are of the form
$| m,i\rangle = \Phi_m(\xperp)f_m^i(z)$.

Hence we can write the sum of the negative eigenvalues as
\beqa\nonumber
\Tr[\Pi_0
[H_A + \varphi]\Pi_0]_- &=& \sum_{m,i}\langle m,i|[\Pi_0
[H_A + \varphi]\Pi_0]|m,i\rangle\\ \label{odt}
&=& \sum_m \left( \sum_i (f_m^i,[-\partial^2_z +
\varphi^{(m)}(z)]f_m^i)\right),
\eeqa
showing that the $f_m^i$'s are the eigenvectors of the
one-dimensional operator $-\partial^2_z + \varphi^{(m)}(z)$.
\end{proof}

\begin{prop}
Let $N,Z,B$ be fixed. Then
\beqa
\label{fub} \Ec^Q(N,Z,B) &\leq& \Ed(N,Z,B) +R_1 + R_2 + R_3,\\
\label{lb} \Ec^Q(N,Z,B) &\geq & \Ed(N,Z,B) - R_1 -
C\int\rho_\psi^{4/3},
\eeqa
with
\beqa\label{secl}
R_1& =&
\left|\sum_{m\in\N_0}\left(\Tr_{L^2(\mathbb{R})}[-\partial^2_z -
\vph]_- - \int\int
[p^{2}-\vph]_-\frac{dpdz}{2\pi}\right)\right|,\\ R_2& =&
D(\rho_\psi - \tilde\rho,\rho_\psi - \tilde\rho),\\ R_3& =&
\sum_{\lambda_N < \lambda_i <\mu(N)} (\lambda_i - \mu(N)),
\eeqa
where $\lambda_i$ and $\tilde \rho $ are defined in the proof.
\end{prop}
\begin{proof}
{\it Upper bound}:\\
First of all we note that for any comparison wave function
$\psi$ and fixed integer $N$
we get
\beq\label{ub}
\Ec^Q \leq (\psi,\Pi_0^NH_N \Pi_0^N\psi) \leq \sum_{i
=1}^{N}(\psi,\Pi_0^N[H_A(\x_i) - Z|\x_i|^{-1}]\Pi_0^N\psi)
+D(\rho_\psi,\rho_\psi),
\eeq
with
\begin{displaymath}
\rho_\psi (\x) = N\sum_{s^{i}}\int |\psi(\x,x_{2},..,x_{N};s^{1},..,s^{N})|^{2}
dx_{2}..dx_{N}.
\end{displaymath}
If we set $\tilde\rho(\x)=\sum_m |\phi_m(\xperp)|^2 \rho^N_m(z)$,
add and subtract $\tilde\rho\ast|\x|^{-1} - \mu(N)$ in the scalar
product and use $\psi = \frac{1}{\sqrt{N!}} \phi_{1} \wedge ... \wedge
\phi_{N}$, where $\phi_i$ is the eigenvector corresponding to the $i$-th
lowest
eigenvalue $\lambda_i$ of the one-particle operator
\beq
\Pi_0 (H_A - Z|\x|^{-1} +\tilde\rho\ast|\x|^{-1} - \mu(N))\Pi_0,
\eeq
as comparison wave function, (\ref{ub}) reads
\beqa\nonumber
\Ec^Q &\leq& \Tr[\Pi_0 (H_A - Z|\x|^{-1} +\tilde\rho\ast|\x|^{-1} -
\mu(N))\Pi_0]_-  - 2D(\rho_\psi,\tilde\rho) \\&&+ D(\rho_\psi,\rho_\psi)
+ N\mu(N)+ \sum_{\lambda_N < \lambda_i <\mu(N)} (\lambda_i -
\mu(N)).
\eeqa
Applying Theorem \ref{tre} and (\ref{Phd}) to the inequality above,
we finally
arrive at the upper bound (\ref{ub}).

{\it Lower bound}:\\
Let $\psi$ denote the minimizer of (\ref{ce}), i.e. $\psi=\psi_{\rm
conf}$. So after again adding and subtracting $\tilde\rho\ast|\x|^{-1} - \mu(N)$
and using the Lieb-Oxford inequality \cite{LO}, we can write the
lower bound on $\Ec^Q$ as follows:
\beqa\nonumber
\Ec^Q &=& (\psi,\Pi_0^NH_N \Pi_0^N)\\ \nonumber &\geq& \sum_{i
=1}^{N}(\psi,\Pi_0^N[H_A(\x_i) -
Z|\x_i|^{-1}+\tilde\rho\ast|\x_i|^{-1} - \mu(N)]\Pi_0^N\psi)\\
\nonumber &&+D(\rho_\psi,\rho_\psi) +N\mu(N) -
2D(\rho_\psi,\tilde\rho) - C\int\rho_\psi^{4/3}\\ \nonumber &\geq&
\Tr[\Pi_0 (H_A - Z|\x|^{-1} +\tilde\rho\ast|\x|^{-1} -
\mu(N))\Pi_0]_- + N\mu\\&& - D(\tilde\rho,\tilde\rho) -
C\int\rho_\psi^{4/3}
\eeqa
Using (\ref{Phd}) we arrive at (\ref{lb}).
\end{proof}
\begin{rem} Due to (\ref{fub}) and (\ref{lb})  the main
contribution to the difference between $\Ec^Q$ and $\Ed$ is given
by
\beq
R_1 =
\left|\sum_{m\in\N_0}\left(\Tr_{L^2(\mathbb{R})}[-\partial^2_z -
\vph]_- - \int\int [p^{2}-\vph]_-\frac{dpdz}{2\pi}\right)\right|,
\eeq
which, by definition, shows that $\Ed$ is the natural
semiclassical approximation of $\Ec^Q$.
\end{rem}
As a corollary, for example by following the way of
\cite{Liebetal1994S} and using coherent states and the above
estimates, or simply by estimating the difference between $\Ed$
and $E^{\rm STF}$ one gets
\begin{cl}
If $Z \to \infty$ with $N/Z$
fixed and $B/Z^{3} \rightarrow 0$, then
\begin{displaymath}
\Ec^{ Q}(N,Z,B)/\Ed(N,Z,B) \rightarrow 1.
\end{displaymath}
\end{cl}

The DSTF theory is equivalent to a three dimensional functional
using  {\it modified} Coulomb potentials,
\beq
V_\chi(\x) = \sum_n \chi^n(\xperp) V_n (z)
\eeq
replacing the attractive Coulomb potential and
\beq
V_{n,m}(z - z')\chi_n(\xperp) \chi_m(\xperp')
\eeq
replacing the Coulomb repulsion, with
\beq
\chi^n(\xperp) =
\left\{ \begin{array}{cc}
1 & {\rm for} \,\, \sqrt{(2n)/B} \leq
|\xperp|\leq \sqrt{2(n+1)/B}\\
0& {\rm otherwise.} \end{array} \right.
\eeq
I.e., for the respective minimizing density of a resulting MSTF
functional, we have $\rho^{\rm MSTF}(\x) = \frac B{2\pi}
\sum_m \rho_m^{\rm DSTF}
(z) \chi_m(\xperp)$ as well as $E^{\rm MSTF} = E^{\rm DSTF}$ for the energy.

Since $V_\chi (0) \sim B^{1/2}$, $V_\chi(\x)$ can be regarded as
a cut off Coulomb potential
\beq
\bar V(\x)= \left \{ \begin{array}{cc} B^{1/2} & {\rm for} \quad
|\x| \leq B^{-1/2},\\  |\x|^{-1} & {\rm for} \quad |\x| \geq
B^{-1/2}.\end{array}\right.
\eeq
Hence, it is obvious that the main contribution to the difference
$\Ed - E^{\rm STF}$ stems from the Coulomb potential in the region
$r\leq B^{-1/2}$,  given by the term
\beq
B\int_{0\leq r \leq B^{-1/2}} |\phi^{\rm STF}|^{3/2} =
O(B^{1/4}Z^{3/2}),
\eeq
which leads to the relation
\beq
\Ed(N,Z,B) - E^{\rm STF}(N,Z,B) = O(B^{1/4}Z^{3/2}).
\eeq
The comparison with Theorem \ref{Ct} shows that the DSTF theory has the same effect as the
introduction of the gradient correction in (\ref{C}) as well as in
(\ref{TS}).

\subsection{The discrete von Weizs\"acker functional}

The variable of the DSTF functional is given by a sum of one
dimensional densities, emphasizing the character of lowest Landau
band particles, whose positions orthogonal to the magnetic field
are \lq\lq frozen" and they therefore only move parallel to the
magnetic field. Now taking into account the result (cf. e.g. Shao
 \cite{Shao1993}) that the first order correction to the semiclassical
description of the one-dimensional free Fermi gas is given by
$-(1/3)\int (\partial_z \sqrt{\rho(z)})^2$, we are motivated to
propose the already mentioned {\it discrete von Weizs\"acker
functional}
\beq
\label{DvW} {\mathcal {E}}^{\rm DW}[\rho] = \sum_{m\in\N_0}\left(-
\frac{1}{3} \int |\partial_{z}\sqrt{\rho_m(z)}|^{2}+  \kappa
\int\rho_m^{3}(z) - Z\int V_m(z)\rho_m(z)\right) + \widetilde
D(\rho,\rho).
\eeq
Since the von Weizs\"acker term appears with negative sign,
(\ref{DvW}) has the same defects as the 
Tomishima-Shinjo functional (\ref{TS}), i.e.
it is not bounded from below and not convex.
Precisely these two features (boundedness and convexity) of the semiclassical 
TF
functionals,
as well the (M)TFW functional, provided not only the existence of a
minimizer but the existence of a solution of the corresponding TF
equation.

As a way out of this problem we can define the energy corresponding
 to (\ref{DvW}) by means of stationary solutions $\rho^N$,
whose variational derivative vanish under the restriction $\sum_m\int
\rho_m =N$ and $\rho \geq 0$. In order to avoid the assumption of
positivity we concentrate on real functions $\psi$, with $\psi^2 =
\rho$ and consider the functional
\beq\label{dwq}
{\mathcal{E}}[\psi] =
 \sum_{m\in\N_0}\left(-
\frac{1}{3} \int |\partial_{z}\psi_m(z)|^{2}+  \kappa
\int\psi_m^{6}(z) - Z\int V_m(z)\psi^{2}_m(z)\right) + \widetilde
D(\psi^{2},\psi^{2}).
\eeq
Let
\beq
D = \{\psi|\ \sum_m\int\psi_m^2 < \infty, \ \sum \int\psi_m^6 <
\infty,\
{\rm and} \
\sum_m \int (\partial_z \psi_m)^2 < \infty\}
\eeq
be the domain of (\ref{dwq}). The question for stationary
points in $D$, under the restriction $\sum_m \int \psi_m^2 = N$, is
equivalent to the existence of a Lagrange parameter $\mu(N)$
and a $\psi^N$, so
that
\begin{equation}
\label {SB}
\left.\frac{d}{dt}\left({\mathcal {E}}[\psi^{N} +  t\eta] +
\mu(N) \int (\psi^{N} +  t\eta)^{2}\right)\right|_{t=0} = 0,
\end{equation}
for each $\eta \in D$. (\ref{SB})
yields the Euler-Lagrange equation, denoting $\psi^N= \psi$,
\beq
\label {ELdvw} (1/3)\partial_{z}^{2}\psi_{m}(z) +
3\kappa\psi_{m}^{5}(z) = [\vph - \mu(N)]\psi_{m}(z)\quad
\forall m \in \N_0.
\eeq
Starting from  (\ref{dwq}), (\ref{ELdvw}) a priori only exists in
the sense of distributions, but if there is a solution for
(\ref{dwq}) then one can conclude that it is even in $C^\infty(\R \setminus \{0\})$.
If there is a solution $\psi^N$ for
$N\in [0,N_c]$, with $N_c \geq Z$,
then we can define the corresponding energy by
\beq
E^{\rm DW}(N,Z,B) = {\mathcal{E}}[\psi^N] = {\mathcal{E}}^{\rm
DW}[(\psi^N)^2].
\eeq

\subsubsection{Recovering the exchange term}

Now we even go a step further. Following the  reflections of Section 1.3.1
concerning the magnitude of the negative von Weizs\"acker term
we guess that $\mathcal{E}^{\rm DW}$ is equivalent to the DDM
functional (\ref{ddmf}) for $B \ll Z^3$. Hence looking at
(\ref{ee}) we suggest another functional,
\beqa\nonumber
&{\mathcal {E}}^{\rm DWHF}[\rho] = \sum_{m\in\N_0}\left(-
(1/3) \int |\partial_{z}\sqrt{\rho_m(z)}|^{2}+  \kappa
\int\rho_m^{3}(z)\right.\qquad\qquad\\ &\qquad\qquad-
\left. Z\int V_m(z)\rho_m(z) - c\ln(B/Z^{4/3})\int \rho_m^2 \right) +
\widetilde
D(\rho,\rho),&
\eeqa
where we recover the exchange energy, which could be compared with the
Thomas-Fermi-Dirac-von Weizs\"acker
functional ([L1]) in the $B=0$ case.

\subsubsection{The one-dimensional DW functional}

In this section we study a toy model obtained by reducing
(\ref{DW}) to a one-dimensional functional and dropping the
Coulomb repulsion, which leads to the functional
\beq
\label{1DW2} {\mathcal {E}}^{\rm 1DW}[\rho] = -
\frac{1}{3} \int |\partial_{z}\sqrt{\rho(z)}|^{2}+  \kappa
\int\rho^{3}(z) - Z\int V_0(z)\rho(z).
\eeq
First of all we consider the corresponding TF functional
\beq
\Eo[\rho] = \ \kappa
\int\rho^{3}(z) - Z\int V_0(z)\rho(z),
\eeq
for which we easily get the following lemma:
\begin{lem}
$\Eo[\rho]$, with $\rho \in L^3(\R)$,
is bounded from below, and there exists a unique minimizing density
$\rho_0 \in L^3$, with $\Eo[\rho_0] = E^{1\rm D}(Z,B)$.
\end{lem}
\begin{proof}
Since $\int V_0 \rho \leq \parallel V_0\parallel_{3/2}\parallel\rho\parallel_3$,
 we get
\beq
\Eo[\rho] \geq \parallel\rho\parallel_3^3 - \parallel V_0\parallel_{3/2}
\parallel\rho\parallel_3.
\eeq
Minimizing over $\parallel\rho\parallel_3$, we get
the first part of the lemma.
The proof of second part works analogously to
Propositions 3.2 and 3.3.
\end{proof}

For the minimizing density $\rho_0$ we get the simple TF equation
\beq
3\kappa\rho_0^2 (z) = ZV_0(z).
\eeq
By the definition (\ref{pot}) one easily sees the relation
\beq\label{sc}
V_0(z) \equiv V_0^B(z) = B^{1/2}V_0^1(B^{1/2}z),
\eeq
which implies an interesting scaling relation for the energy
$E^{1\rm D}(Z,B)$:
\begin{lem}
\beq\label{scalrel}
E^{1\rm D}(Z,B) = Z^{3/2}B^{1/4}E^{1\rm D}(1,1).
\eeq
\end{lem}
\begin{proof}
Using the scaling relation (\ref{sc}) and defining
\beq
\rho(z) = B^{1/4}Z^{1/2}\bar \rho(B^{1/2}z),
\eeq
we get
\beq
\Eo[\rho] = B^{1/4}Z^{3/2}{\mathcal E}^{1\rm D}_{1,1}[\bar \rho].
\eeq
\end{proof}

In the next Theorem we  point out that for $B \ll Z^2$, the energy
(\ref{scalrel}) is the semiclassical approximation of $\Tr
[-\partial_z^2 - ZV_0(z)]_- $, the sum of all negative eigenvalues
of $-\partial_z^2 - ZV_0(z)$.

\begin{thm}\label{corproof}
Let $B \leq Z^2$ and $\psi \in C_0^\infty(B(0,B^{-1/2}))$.
Then
\beq\label{cor}
\Tr(\psi[-\partial_z^2 - ZV_0(z)]_- ) = Z^{3/2}B^{1/4}\int
\int\frac{dzdp}{2\pi} \psi [p^2-V^1_0(z)]_- -O(B^{3/4}Z^{1/2}),
\eeq
and there is a constant $C$, such that
\beq
|\Tr[-\partial_z^2 - ZV_0(z)]_-  - Z^{3/2}B^{1/4}E^{1\rm D}(1,1)|
\leq CB^{3/4}Z^{1/2}.
\eeq
\end{thm}

\begin{proof}

Let us rewrite
\beq
-\partial_z^2 - ZV^B_0(z) = B^{1/2}Z[-(B^{1/2}Z)^{-1}
\partial_z^2 - V_0^1(B^{1/2}z)]
\eeq
and define the unitary operator
\beq
(U(l)\psi)(z) = l^{1/2}\psi(l z).
\eeq
With $l=B^{-1/2}$
and the fact that unitary transformations do not change traces
we derive
\beq\label{b1}
\Tr [-\partial_z^2 - ZV^B_0(z)]_- = B^{1/2}Z\Tr [- h^2\partial_z^2
- V_0^1(z)]_-,
\eeq
where
\beq\label{b2}
h = B^{1/4}Z^{-1/2}.
\eeq
So we have the self-adjoint Schr\"odinger operator
\beq
H = -h^2\partial_z^2 - V_0^1(z),
\eeq
and its symbol
\beq
h(z,p) = p^2 - V_0^1(z).
\eeq
Using the relation
\beq
|\partial^\nu V_0^1(z)| \leq C^\nu \,\,\, {\rm on} \,\,\,
B(0,2) = \{z|\, |z| < 2\}
\eeq
and by means of  some cut off function $\psi \in C_0^\infty$,
\cite{IS} Theorem 6.1 yields
\beq\label{e1}
\Tr(\psi(z)[-h^2\partial_z^2 - V^1_0(z)]_-)
= h^{-1}\int \int\frac{dp dz}{2\pi}\psi(z)[p^2 - V_0^1(z)]_-
- O(h),
\eeq
which together with (\ref{b1}) and (\ref{b2})
immediately implies (\ref{cor}).

\noindent
Recall, that by $[t]_- = t $ for $t\leq 0$ and $0$
otherwise.

In order to tackle the outer zone $\{ z| 1 \leq |z| \leq \infty\}$,
we note that the potential fulfills
\beq
|\partial^\nu V_0^1 (z)| \leq C^\nu |z|^{-\nu} \quad {\rm for} \quad
|z| \geq 1.
\eeq
So by definition of the scaling functions $l(z)=|z|$ and $f(z) = 1$,
we can use \cite{IS} Theorem 7.1, which states that with a cut off function
$\psi \in C^\infty(B(1,\infty))$, we get
\beq\label{e2}
\left|\Tr(\psi[-h^2\partial_z^2 - V^1_0(z)]_-) - h^{-1}\int
\int\frac{dp dz}{2\pi}\psi[p^2 - V_0^1(z)]_-\right| \leq
h\int_1^\infty dz l^{-2} \leq h.
\eeq
Combining (\ref{e1}) and (\ref{e2}) together with (\ref{b1}) and (\ref{b2})
completes the proof of Theorem \ref{corproof}.
\end{proof}

Turning back to the functional (\ref{1DW2}), we recall that the
corresponding energy
has to be defined by means of the solution of the TF equation
\beq\label{equ}
\partial_z^2 \psi(z) + 3\kappa \psi^5 (z) - ZV_0^B (z)\psi(z) = 0.
\eeq
Introducing the scaling relations
\beq
\psi(z) = B^{1/8}Z^{1/4}\varphi(zB^{1/2}), \quad z \to zB^{1/2},
\eeq
(\ref{equ})
can be written as
\beq\label{equ2}
\mbox{$\frac{B^{1/2}}{Z}$} \partial_z^2 \varphi(z) + 3\kappa\varphi^5 (z) -
V_0^1 (z) = 0.
\eeq
We set
$\eps \equiv B^{1/2}/Z \ll 1$ and make the ansatz $\varphi = \varphi_0 +
\eps\varphi_1$
to get an approximate solution of (\ref{equ2}).
Rescaling and inserting into (\ref{1DW2})
yields
\beq
E^{1\rm DW} = {\mathcal {E}}^{1\rm DW}[\psi^2] = Z^{3/2}B^{1/4}E^{1\rm D}(1,1)
- O(B^{3/4}Z^{1/2}),
\eeq
which is in accordance with (\ref{cor}) and justifies a posteriori
the introduction of the negative von Weizs\"acker term in
(\ref{DW}) and (\ref{1DW2}).

\section{Some concluding remarks}

Throughout this paper we have studied, among others, two functionals, the MTFW,
which represents an approximation to the  full quantum mechanical
energy $E^Q$, and the DW functional, which should approximate the
ground state energy $\Ec^Q$ of particles in the lowest Landau
band. Now we can ask for the magnitude of $B$, for which the use of
DW becomes more reasonable than MTFW. This question is connected
with the estimation of
\beq\label{qqc}
\left|E^{\rm Q}(N,Z,B) - E^{\rm Q}_{\rm conf}(N,Z,B)\right|.
\eeq
For $B \ll Z^3$ this can be compared with the difference of the corresponding
semiclassical approximations
\beqa\nonumber
&\left|B\int |\phi^{\rm MTF}(x)|_{+}^{3/2}dx + 2B\sum_{i\geq
1}\int|\phi^{\rm MTF}(x) - 2iB|_{+}^{3/2}dx - B\int |\phi^{\rm
STF}(x)|_{+}^{3/2}dx\right|& \\ \label {012} & \leq C \int
|Z|x|^{-1} - 2B|_{+}^{3/2}dx \leq CZ^{3}/B^{1/2}.&
\eeqa
From the preceding sections we guess, on the one hand,
\beq
|E^{\rm DW} - \Ec^Q| \leq o(B^{4/5}Z^{3/5}),
\eeq
and on the other hand we know
\beq
|E^{\rm MTFW} - E^Q| \leq O(Z^2) + O(B^{4/5}Z^{3/5}).
\eeq
Hence, one might expect the DW theory to give  the better description
of the quantum mechanical energy, if $Z^{3}/B^{1/2}\leq
B^{4/5}Z^{3/5}$,
or, in other words, $B \geq Z^{24/13}$.

Summing up, we  suggest to use MTFW theory
for $B \leq Z^{24/13}$ and DW for $Z^{24/13} \leq B \ll Z^{3}$.\\

\noindent
{\bf Acknowledgement.}
The author is very thankful to Jakob Yngvason for proofreading and 
many helpful discussions and he furthermore thanks Robert Seiringer
for many valuable comments.

\end{document}